\documentclass[sigconf,screen,nonacm]{acmart}
\AtBeginDocument{%
  }
\usepackage{xspace}
\usepackage{enumitem} 
\usepackage{booktabs}
\usepackage{graphicx}
\usepackage{svg}
\usepackage{float}
\usepackage{booktabs,tabularx,array}
\usepackage[utf8]{inputenc}
\usepackage[T1]{fontenc}
\usepackage{xcolor}
\usepackage{tikz}
\usepackage{soul}
\usepackage{multirow}
\usepackage[utf8]{inputenc}

\usepackage{fvextra} % fancyvrb
\DefineVerbatimEnvironment{prompt}{Verbatim}{
  breaklines=true,
  breakanywhere=true,
  breaksymbolleft=\tiny\(\hookrightarrow\)\,,
  numbers=left, numbersep=6pt,
  frame=single,
  fontsize=\small,
  obeytabs=true, tabsize=2
}

\newcommand{\codeex}[1]{\texttt{\footnotesize\detokenize{#1}}}
\setcopyright{acmlicensed}
\copyrightyear{2018}
\acmYear{2018}
\acmDOI{XXXXXXX.XXXXXXX}
\acmConference[Conference acronym 'XX]{Make sure to enter the correct
  conference title from your rights confirmation email}{June 03--05,
  2018}{Woodstock, NY}
\acmISBN{978-1-4503-XXXX-X/2018/06}

\begin{document}

\title{Disclose with Care: Designing Privacy Controls in Interview Chatbots}

% \affiliation{%

\author{Ziwen Li}
\email{li.ziw@northeastern.edu}
\affiliation{%
  \institution{Northeastern University}
  \city{Boston}
  \state{Massachusetts}
  \country{USA}
}

\author{Ziang Xiao}
\email{ziang.xiao@jhu.edu}
\affiliation{%
  \institution{Johns Hopkins University}
  \city{Baltimore}
  \state{Maryland}
  \country{USA}
}
\author{Tianshi Li}
\email{tia.li@northeastern.edu}
\affiliation{%
  \institution{Northeastern University}
  \city{Boston}
  \state{Massachusetts}
  \country{USA}
}

\renewcommand{\shortauthors}{Li et.al.}

\newcommand\tldr[1]{\textcolor{gray}{\textit{[#1]} \\}}
\newcommand{\system}{\emph{EDIT}\xspace}
\newcommand{\tianshi}[1]{\noindent{\textcolor{blue}{\textbf{\#\#\# Tianshi:}{#1} \#\#\#}}}
\newcommand{\zx}[1]{\noindent{\textcolor{orange}{\textbf{\#\#\# Ziang:}{#1} \#\#\#}}}

\newcommand{\aaron}[1]{\noindent{\textcolor{orange}{\textbf{\#\#\# Aaron:}{#1} \#\#\#}}}

\newcommand{\red}[1]{#1}

\renewcommand{\sectionautorefname}{Section}
\renewcommand{\subsectionautorefname}{Section}
\renewcommand{\subsubsectionautorefname}{Section}

\definecolor{pinkcustom}{HTML}{E24A70}
\definecolor{bluecustom}{HTML}{4A90E2}
\definecolor{orangecustom}{HTML}{F5A623}
\definecolor{greencustom}{HTML}{7ED321}

\begin{abstract}
Collecting data on sensitive topics remains challenging in HCI, as participants often withhold information due to privacy concerns and social desirability bias.
While chatbots' perceived anonymity may reduce these barriers, research paradoxically suggests people tend to over-share personal or sensitive information with chatbots.
In this work, we explore privacy controls in chatbot interviews to address this problem. The privacy control allows participants to revise their transcripts at the end of the interview, featuring two design variants: free editing and AI-aided editing.
In a between-subjects study \red{($N=188$)}, we compared no-editing, free-editing, and AI-aided editing conditions in a chatbot-based interview on a sensitive topic.
Our results confirm the prevalent issue of oversharing in chatbot-based interviews and show that AI-aided editing serves as an effective privacy-control mechanism, reducing PII disclosure while maintaining data quality and user engagement, thereby offering a promising approach to balancing ethical practice and data quality in such interviews.

\end{abstract}

\begin{teaserfigure}
  \centering
  \includegraphics[width=.8\linewidth]{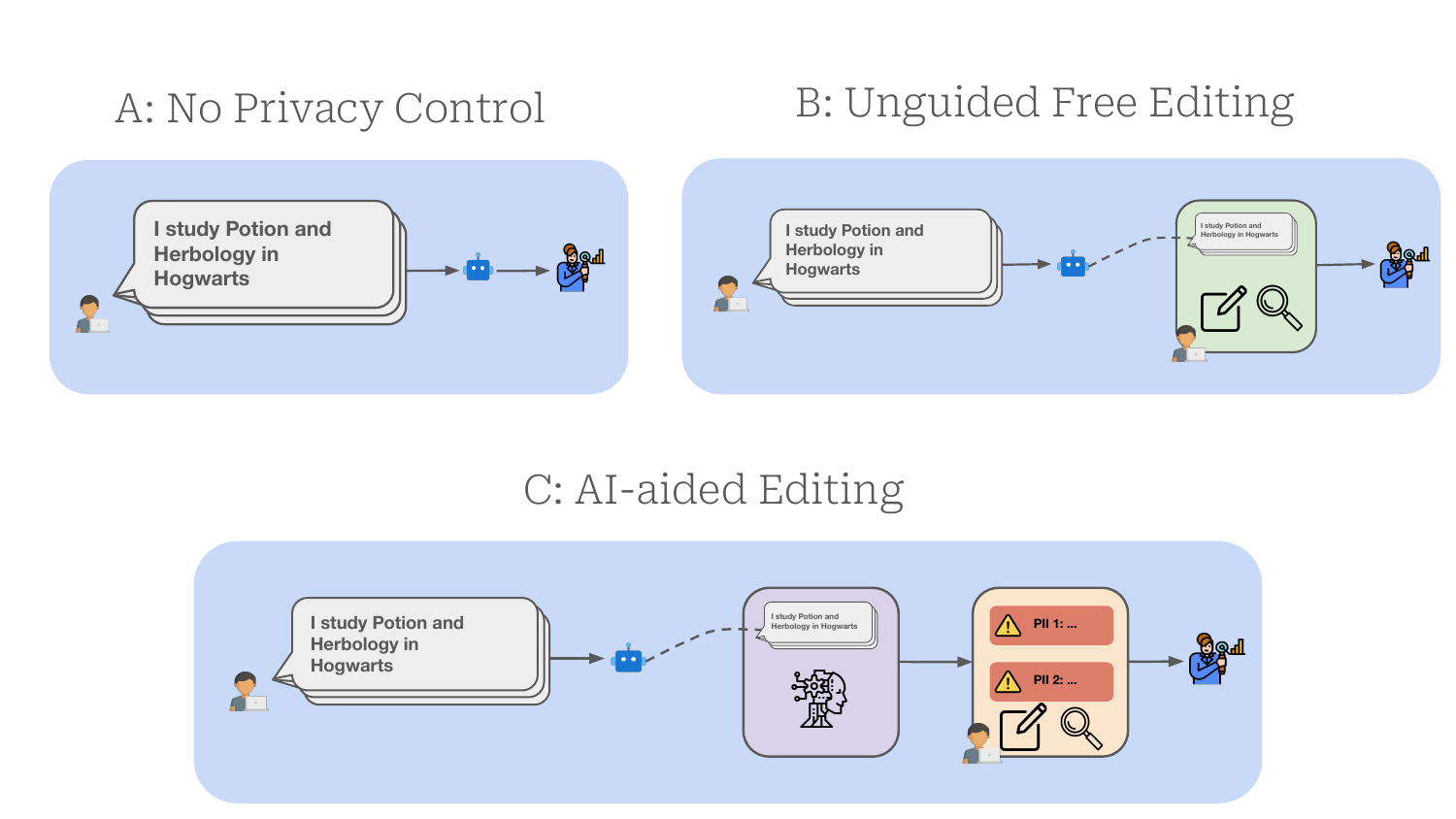}
  \caption{We examine three experimental conditions for studying privacy controls for interview chatbots about their impact on privacy protection and data quality. \textbf{(A) No Privacy Control (Baseline):} Chatbots forward participants’ raw responses directly to researchers without intervention, preserving data authenticity but offering no safeguards. \textbf{(B) Unguided Free Editing:} After the interview, participants are prompted to freely review and edit their conversation logs, enabling autonomy yet without structured guidance. \textbf{(C) AI-aided Editing:} Large language models (LLMs) automatically detect and flag potential personally identifiable information (PII), supporting participants with guided post-interview editing to mitigate privacy leakage while retaining user control.}
  \label{fig:teaser}
\end{teaserfigure}





\maketitle

\section{Introduction}
In human-computer interaction (HCI) and social science research, data quality on sensitive topics is often compromised by participants' tendency to withhold or misrepresent information \cite{10.1145/3613904.3642732,panicker2024forms}.
This phenomenon, often driven by concerns about privacy \cite{gross2020validity}, social desirability bias \cite{latkin2017relationship,Caputo2017SocialDB}, or potential repercussions \cite{stopczynski2014privacy}, directly compromises the validity and reliability of self-report data. When individuals obscure their true opinions, behaviors, or experiences on sensitive topics, such as health conditions, financial practices, or personal beliefs, researchers are left with an incomplete or inaccurate understanding of the phenomena under investigation \cite{NIH_DataIntegrity,NIH_SocialDesirability}. 
Consequently, such studies yield biased insights that produce flawed theoretical models, ineffective design recommendations, and misguided interventions, failing to address societal challenges and cause harm.

To address these data quality challenges, chatbots demonstrate considerable potential as scalable interviewers, offering interactive and conversational experiences conducive to collecting data from participants compared to paper-based surveys and face-to-face interviews, particularly concerning sensitive topics \cite{lind2013survey,locke1992computer,richman1999meta,weisband1996self,lucas2014s}.
This is further supported by findings indicating that individuals often disclose more sensitive information to computer-assisted interfaces due to perceived non-judgment, reduced social inhibition, and confidentiality assurance \cite{PICKARD201623,cassell2001embodied}.
However, this affordance also carries the risk of oversharing~\cite{xiao2020tell}, as interviewees may disclose more sensitive or identifiable information than is necessary for the questions, or more than they would naturally share without using a chatbot.
This problem is further exacerbated as LLM-powered chatbots exhibit more human-like behavior, which can induce even greater levels of disclosure~\cite{zhan2025malicious, zhang2024s}.  A trending industrial example is Anthropic Interviewer, an AI-powered tool that can scale up interviews with high quality \cite{handa2025anthropicinterviewer}.
However, the rich details elicited by the chatbot also result in high vulnerability of privacy leakage, due to the lowered barrier and ease of re-identifying interviewees with web-search LLM agents \cite{li2026agentic}.
Such unnecessary disclosures may place users at risk of privacy violations or identity theft, particularly if a data breach occurs or the data is released without sufficient privacy-preserving measures.

Common practices to address oversharing and the resulting privacy risks include obtaining informed consent before data collection, ensuring and communicating secure storage procedures to reduce breach risks and build trust, and enforcing strict access controls and data minimization once the data has been uploaded~\cite{xiao2020tell}.
These methods offer essential safeguards by fostering transparency and accountability in the research process.
However, they remain insufficient to fully protect users from unwanted or impulsive disclosures, given the inherent power asymmetry between researchers and study participants.
The efficacy of informed consent has long been questioned~\cite{herz1992informed}, complicated by participants’ differing levels of literacy and understanding of research information~\cite{kang2015my}. Furthermore, the question of how much detail is sufficient in an informed consent remains under debate~\cite{stanley1998informed}.

Accordingly, endowing users with more controls, such as controlling what information to share~\cite{araujo2022osd2f} and how the shared information is used~\cite{xiao2020tell}, has been proposed as a way to repair this power asymmetry.
Nevertheless, this also raises researchers’ concerns about the authenticity of participants’ responses~\cite{xiao2020tell}.
Currently, there is limited research on this topic, which remains an essential gap in achieving a balance between ethical research practices and data validity.
In our research, we aim to bridge this gap by presenting the first study to examine privacy controls in chatbot-based interviews, guided by two research questions: (1) \textit{How would participants use privacy controls in chatbot-based interviews}, and (2) \textit{How would privacy controls impact participant privacy, data quality, and user engagement in chatbot-based interviews?}

To this end, we developed an interview chatbot system, featuring two modes of post-interview privacy controls: free editing and AI-aided editing. In both modes, participants are provided with an opportunity to freely revise their responses after the interview and are informed of this opportunity beforehand.
The AI-aided editing mode is motivated by the recurring theme in usable privacy literature of limited privacy awareness and need for scaffolds to support rational privacy decisions~\cite{zhou2025rescriber, kang2015my, zhang2024s}.
Our AI-aided mode scans the entire interview log for personally identifiable information (PII), and generates suggestions to either redact the PII to a type placeholder (e.g., [NAME1]) or abstract them to a more general level (e.g., replacing ``18 years old'' with ``young adult''), which participants can optionally accept and apply.
Crucially, we aim to leverage AI as a gentle reminder to address gaps in awareness and privacy literacy, rather than as a coercive or overly directive intervention. In doing so, our system aims to preserve user autonomy, restore user control, and reduce the power asymmetry.

To empirically evaluate the behavioral differences stemming from these privacy controls, we conducted a between-subjects study ($N=188$) comprising three conditions: free editing, AI-aided editing, and no editing (the control group).
We chose to design the structured chatbot-based interview focusing on the sensitive topic of \textit{using AI for job interviews} for its contemporary relevance, inherent sensitivity~\cite{zhang2025secret, sarkar2025ai}, and accessibility to a pool of qualified participants.
Note that in this work, we focus our analysis on evaluating the privacy control methods rather than the interview topic itself.
Conversation log data, both pre- and post-editing, were collected to analyze the effect of the privacy controls, specifically on privacy preservation, interview data quality, and participant engagement.

Our study shows that post-hoc privacy controls can convert oversharing into small, focused edits without diminishing the substance of what participants say. The AI-aided privacy control significantly improved privacy protection, measured as PII reduction, by steering users toward targeted replacements and abstractions of identifiable details (e.g., names, times, affiliations, places) rather than blunt deletions or unrecognizable rewrites. The privacy gain came with no significant cost to interview utility, as responses remained comparably relevant and clear, engagement levels stayed stable, and editing effort was light and focused. Whereas free editing alone sometimes led to elaboration and new disclosures, AI suggestions made risk visible and safe actions easy.
We also observed a perception-behavior gap: although participants across conditions reported similar senses of privacy control and confidence in disclosure, AI-aided users enacted more protective edits, highlighting the value of visible, reversible, span-level cues. 
By foregrounding transparency, maintaining user agency, and embedding privacy safeguards into the flow of data collection, our study offers a practical and ethically aligned response to the long-standing challenges in subjective data collection on sensitive topics.

In summary, our contributions are the following:
\begin{itemize}
    \item We designed and implemented a privacy-aware chatbot interface that combines structured questioning with conversational flexibility and incorporates two privacy control features, offering a novel approach to eliciting sensitive disclosures while preserving participant autonomy and privacy.

    \item We conducted a controlled experiment to evaluate the two proposed post-hoc privacy control features, using comparative analyses of conversation logs to assess effects on privacy preservation, data quality, and participant engagement.

    \item We offer design guidelines on the integration of privacy-preserving mechanisms into chatbot-based interview studies, enabling ethically-aligned,  scalable data collection across diverse domains of HCI and social research.
\end{itemize}

\section{Related Works}
\subsection{Challenges in Data Collection for Research on Sensitive Topics}
Prior work has demonstrated that obtaining accurate self-reports on sensitive topics, including health, finance, and ethical dilemmas, is challenging. Participants often underreport or distort their responses due to social desirability bias \cite{NIH_SocialDesirability, Caputo2017SocialDB,tourangeau2007sensitive, paulhus1991responsebias}, and privacy concerns further deter disclosure \cite{acquisti2015privacy}. Similar patterns have been observed in online communities and digital platforms, where disclosure is influenced by contextual norms and perceived surveillance \cite{madejski2012failure, boyd2014networked, stutzman2013silent}. Additionally, research on social media has highlighted the fragility of privacy decision-making, often stemming from regrets over disclosure \cite{wang2013privacy} and underestimation of audience size \cite{10.1145/3706598.3714074}. Recent studies have further explored this through the lens of cognitive styles and automated systems, showing that risk-benefit evaluations drive information sharing, while user-friendly interfaces can support better privacy behaviors \cite{BARTH20171038}.

Despite voicing strong privacy concerns, users frequently disclose personal information when they trust the recipient, revealing a disconnect between attitudes and behaviors \cite{BARTH20171038}. This phenomenon, known as the privacy paradox, describes how individuals' actions do not align with their stated concerns, often justified by default settings or perceived invulnerability \cite{nicol2022revealing}.
Research has unraveled this paradox by examining factors like cognitive and psychological influences, technological aspects, and sociocultural elements, emphasizing the role of perceived benefits outweighing risks in digital marketing and e-commerce contexts \cite{wang2025unraveling}. These issues underscore the entanglement of self-report data collection with impression management and privacy concerns, prompting the development of alternatives that reduce judgment and build trust. 

Accordingly, HCI research increasingly focuses on designing data collection methods that address disclosure hesitation and balance impression management with privacy protection, especially for stigmatized topics \cite{lee2020hear, 10.1145/3274288}.
Our study advances this research by emphasizing post-hoc privacy mechanisms that act after disclosure to address the over-sharing of privacy information.
Notably, we observe a discrepancy between the subjective perception in privacy disclosure and the objective detected PII which echos the classic privacy paradox issues, and demonstrate the efficacy of AI-aided editing in addressing the gaps in privacy awareness and promoting more responsible sensitive data sharing behaviors.

\subsection{Chatbots for Sensitive Interviews}
Conversational agents have been investigated as means to encourage self-disclosure in sensitive areas. Prior studies indicate that virtual agents promote greater openness than human interviewers, as they are seen as less judgmental and more approachable \cite{gratch2014s, lee2020hear}. Recent research has applied this to health and well-being, with chatbots aiding disclosure in mental health and stress management \cite{10.1145/3588967.3588971, fitzpatrick2017woebot}. Further work has examined trust, empathy, and disclosure in chatbot interactions, highlighting opportunities and risks \cite{mireshghallah2024trustbotdiscoveringpersonal}. Emerging evaluations of generative AI chatbots for mental health reveal high user engagement and positive impacts, such as improved relationships and trauma healing, while stressing the need for enhanced safety features and human-like capabilities \cite{siddals2024experiences}. Meanwhile, scholars caution that power imbalances in chatbot-mediated consent can mirror wider issues of autonomy and control \cite{xiao2023informed, romare2025towards}, reflecting broader concerns about expectation gaps in conversational agents \cite{luger2016limitations}. Collectively, these findings establish chatbots as promising but risky mediators for sensitive interviews, necessitating a balance between disclosure benefits and risks of overexposure and loss of control.

Our proposed chatbot system tackles these issues by integrating a post-interview privacy leakage detection module that identifies sensitive information without interrupting the conversation, addressing the risk of overexposure. The inclusion of a manual editing interface and an optional AI-aided-editing feature empowers users to retain control over their disclosures, mitigating power asymmetries and enhancing trust, as suggested by prior work on autonomy and consent. By conducting a controlled user study comparing free editing and AI-aided editing against a control group, we empirically validated how these mechanisms improve disclosure quality and reduce the expectation gaps identified in previous chatbot research, offering a scalable solution for sensitive interview contexts.

\subsection{Limitations of Current Privacy Consent Approaches}
Static consent models have been criticized for failing to provide meaningful user control \cite{solove2013consentdilemma, bietti2020consentfreepass}. Alternative approaches, such as dynamic consent and privacy nudges, demonstrate methods to integrate ongoing, context-aware decision-making into digital systems \cite{kaye2015dynamic, schaub2015notices}. Research on behavioral advertising and location sharing illustrates that disclosure preferences are highly contextual and difficult to generalize \cite{consolvo2005location, ur2012smart}. These designs frequently face challenges in balancing privacy protection with usability, especially in real-time interactions \cite{acquisti2017nudges, shilton2012participatory, jiang2024selfdisclosureaiparadoxtrust}. In AI and chatbot settings, automated data sanitization techniques mitigate leakage risks \cite{carlini2021extraction, presidio} but may disempower users by reducing their autonomy to decide data to sanitize. Recent analyses highlight the evolving landscape, where frameworks like the EU AI Act demand risk-based governance for high-risk systems, emphasizing transparency and human oversight to address gaps in traditional consent models \cite{cloudsecurityalliance2025ai}. Thus, prior work highlights a key gap: integrating automation with user-directed privacy control in an effective and autonomy-respecting way. To address this, consent should be reconceptualized as an ongoing process throughout the interaction lifecycle, with special focus on the timing of disclosure in conversational systems.

Our work responds to these limitations by introducing a privacy-aware chatbot that postpones privacy interventions until after the interview, complementing the mainstream pre-sharing consent models. The system’s leakage detection module and dual editing options, manual and AI-aided, provide users with flexible, post-hoc privacy control, addressing the need for context-aware decision-making without sacrificing autonomy. By evaluating these features in a user study on AI use in job interviews, we aim to demonstrate how our approach enhances transparency and user agency, aligning with the EU AI Act’s emphasis on human oversight, and offers a practical solution to the persistent gap between automation and user-led privacy control.

\subsection{Designing for Autonomy and Privacy in Chatbot Systems}
Growing interest in autonomy-supportive design stems from research identifying strategies like scaffolding, choice provision, and just-in-time interventions to enhance digital autonomy \cite{wang2023authonomy}. Prior studies on informed consent show that conversational interfaces can improve user comprehension and mitigate power imbalances \cite{xiao2023informed}. Moreover, privacy profile autonomy demonstrates how systems can be customized to individual preferences while maintaining scalability \cite{romare2025towards}. Related work emphasizes keeping humans in the loop in privacy-aware AI system design \cite{wang2021cass, 10.1145/3637330} and addresses trust, transparency, and autonomy in AI-assisted decision-making \cite{lai2022humanaicollaborationconditionaldelegation, 10.1145/3613905.3638184}. Recent literature reviews on user privacy concerns in conversational chatbots adopt a social informatics approach, identifying themes like data handling, consent, and ethical implications to inform HCI design practices \cite{li2024literature}. Drawing on these insights, our work incorporates post-hoc, optional privacy reminders in chatbot interactions, expanding autonomy-supportive mechanisms to sensitive self-disclosure contexts.

This approach aligns with the wider HCI trend toward privacy-by-design that prioritizes user agency, illustrating how lightweight interventions can bolster autonomy without compromising the usability or efficacy of conversational systems. Our system builds on these principles by integrating a customizable chatbot interface and post-interview editing options, directly addressing the need for human-in-the-loop design. 
\section{\system: Enhanced Disclosure Intervention Tool}

\begin{figure*}[h!]
  \centering
  
  \includegraphics[
    width=.7\textwidth,
    height=0.9\textheight,
    keepaspectratio
  ]{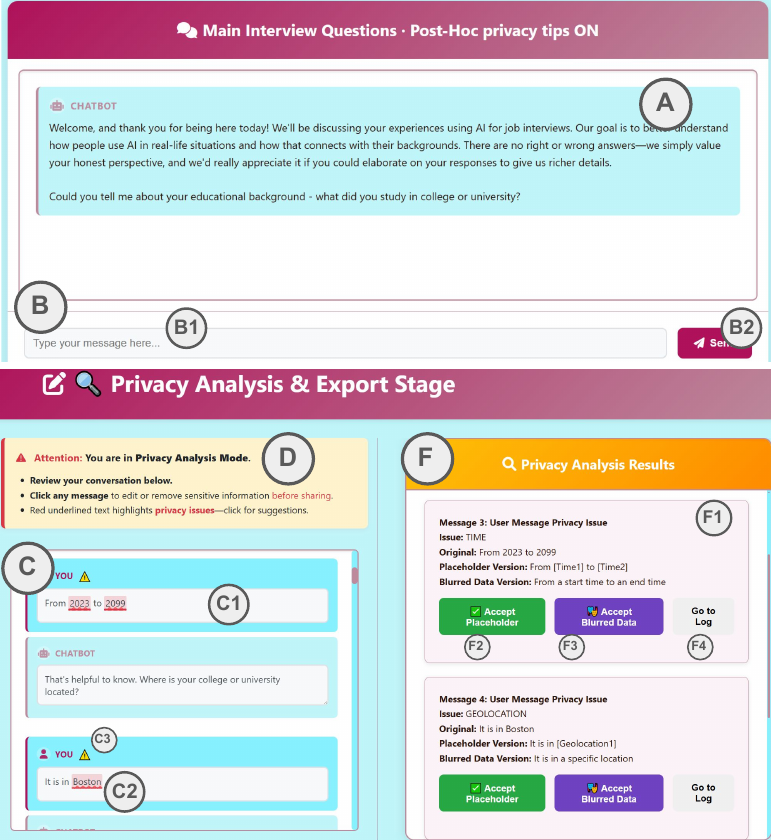}

  \caption{A snapshot of \system user experience in AI-aided Editing during an interview about \textit{using AI for job interviews}. The interface starts with a chatbot interface (Top) which includes a scrollable interface to demonstrate the chat logs between users and the interview chatbot (A) and the text area (B) that contains the text box (B1) and the send button (B2). The interface in the privacy control phase (Bottom) consists of a conspicuous prompt (D) to notify participants about the following steps, a free-editing area (C) with interview logs (C1) that enable the in-text highlights of PIIs (C2), and a button in warning signs that redirect participants to the corresponding privacy issue card by clicking it (C3). The interface flags potentially sensitive text with a red wavy underline and a privacy tooltip (F) including privacy issue cards (F1) that explain the risk and offer three edit buttons: replace the text with placeholders (F2) or blur sensitive details (F3), with an ignore control if users want to revert to their original logs (F4). For the free-editing group, the participants will only get access to (C) and (D) while the participants in Control group will not do post-interview editing.}
  \label{fig:snapshot}
\end{figure*}

We developed a web-based application, hereafter referred to as the \textit{Enhanced Disclosure Intervention Tool} (\system), to conduct semi-structured, multi-turn interviews with online participants, and then implemented the two proposed privacy control features based on it. \autoref{fig:snapshot} depicts the proposed extension, illustrating example actions and outcomes using \system.

\subsection{Base Chatbot System}

\begin{figure}[h!]
  \centering
  
  \includegraphics[
    width=0.8\columnwidth,
    height=0.9\textheight,
    keepaspectratio
  ]{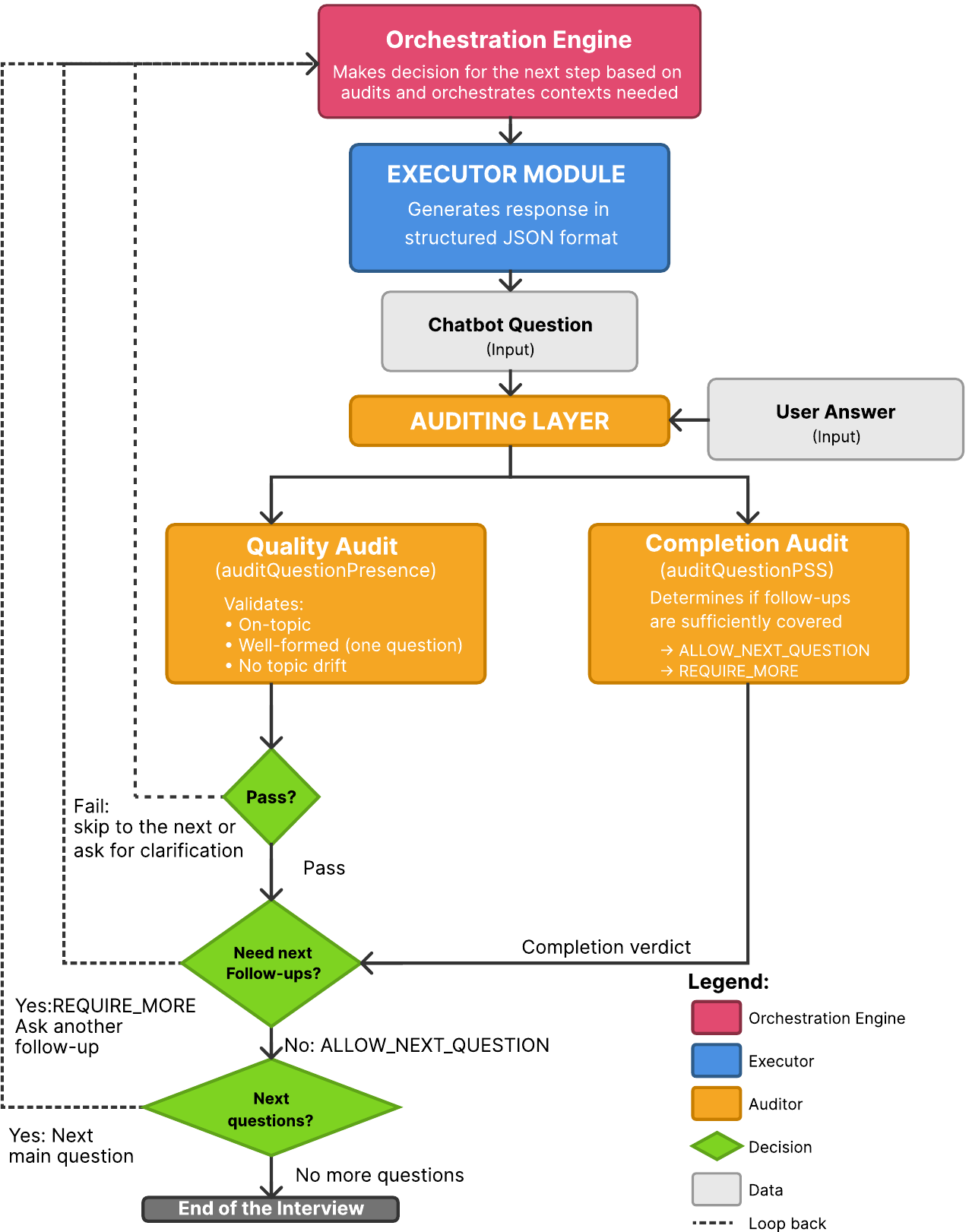}
  
  % }
  \caption{A diagram for conversational orchestration visualization. The process begins when \textcolor{pinkcustom}{the orchestration engine} selects the next main question and prompts \textcolor{bluecustom}{the Executor} to generate a structured JSON response, which is delivered to the user as the chatbot’s question. Once the user answers, \textcolor{orangecustom}{the Auditing Layer} evaluates the exchange through two LLM-driven audits: \textcolor{orangecustom}{a quality audit} ensuring the response is on-topic, well-formed, and free of topic drift, and \textcolor{orangecustom}{a completion audit} determining whether sufficient follow-ups have been covered. Based on these verdicts, \textcolor{pinkcustom}{the orchestrator} either triggers an additional follow-up, asks for clarification, or advances to the next main question. This loop continues until no further questions remain, concluding the interview.}
  \label{fig:orchestration}
\end{figure}

We implemented a chatbot with multi-component orchestration engine that manages the conversational flow as the base system for interviews. The orchestration is shown in \autoref{fig:orchestration}.
Recent works on dialog agent suggests that prompting a frozen LLM to generate orchestration signals can facilitates the process of deciding what actions the dialogue agent should take to effectively to achieve specific goals during the dynamic interactions with users \cite{fu2023improving,deng2023plug,li2023camel}. Inspired by those work, we similarly adopt an orchestrator-executor design to ensure smooth, consistent interview experiences throughout the chatbot session. This engine operates in a loop, processing each user message through a series of specialized LLM-powered modules to generate a contextually appropriate and protocol-adherent response and a smooth transition between follow-ups and the main questions in the designed semi-structured interview (see the detailed prompts in \autoref{sec: prompts}). We implemented an Executor-Auditor state-machine loop because in early pilot prototyping, a single-pass generator occasionally drifted and skipped required follow-ups, barely finishing a single interview. Adding lightweight quality and completion audits provided more reliable adherence, reaching 100\% completion rate and topic coverage in our pilot testing.

\subsubsection{The Orchestration Engine.}

The orchestration engine is responsible to translate the audit results from last step to the executable prompt that can be directly fed into the executor for the next generation. It controls the smooth transition of the state machine. For different tasks such as asking another follow-up or proceeding to the next main question, the orchestration engine would fetch the context needed accordingly.

\subsubsection{The Executor Module.}

The primary conversational turn is generated by the Executor module. After receiving a user's message, the orchestrator constructs a prompt for the Executor that includes the current main question, the conversation history, and a dynamically generated list of ALLOWED\_ACTIONS (e.g., ASK\_FOLLOWUP, NEXT\_QUESTION). The Executor is tasked with generating a response in a structured JSON format that specifies which action to take and the corresponding utterance. The system prompt for this module explicitly instructs it to ``elicit a concrete personal story'' while maintaining a concise and conversational tone. This is because high-quality interview data requires interviewees to describe their actual experiences and feelings, rather than speaking in high-level or abstract terms. Participants could still choose to say they do not have related experiences and skip any question.

\subsubsection{The Auditing Layer.}

To ensure adherence to the research protocol, the Executor's output and the user's input are evaluated by an Auditing Layer composed of two distinct LLM-driven auditors.

\textbf{Completion Audit:} The \textit{auditQuestionPSS} function serves as the primary mechanism for advancing the interview. After a user provides a response, this auditor is invoked to determine if the conversation has sufficiently covered the predefined follow-ups associated with the current main question. It is prompted with the conversation history and a JSON list of required follow-up topics. Its verdict, either \textit{ALLOW\_NEXT\_QUESTION} or \textit{REQUIRE\_MORE}, is used by the orchestrator to decide whether to transition to the next main question or to ask another follow-up.

\textbf{Quality Audit:} To maintain conversational quality, the \textit{auditQuestionPresence} function acts as a quality control check on the Executor's generated utterance. It validates that the bot's message is on-topic, well-formed (e.g., contains exactly one question), and does not prematurely drift to other topics. If the audit fails, it can trigger a regeneration of the bot's response.

\subsection{\system Privacy Control and User Experience Design}

Apart from the base chatbot system, the core of \system is the AI-aided privacy controls extension that can be attached to similar interview chatbot systems. To investigate the effects of different privacy controls, \system is designed with three distinct experimental modes. 

\textbf{Free Editing Mode:}  After the interview concludes, the user is presented with a fully editable log of the entire conversation. They can freely add, remove, or modify any part of their or the chatbot's messages before final submission. Red wavy underlines feature and PII detectors are not provided in the Free Editing Mode. 
    Participants are informed of the opportunity to make post-hoc edits by the initial consent form, the pop-up window after the interview, and the informative prompt on the chatbot interface throughout the interview process.

\textbf{AI-aided Editing Mode:} In addition to the free editing capabilities, this mode offers AI-assisted, granular privacy controls. After the interview session, the system runs the \textit{detectPrivacyWithAI} module on the conversation log. This module uses an LLM to identify potential PII according to a predefined taxonomy from \citet{zhou2025rescriber} (e.g., NAME, AFFILIATION, HEALTH\_INFORMATION). The full prompts are listed in \autoref{sec: prompts}.
The user is then shown the log with highlighted PII alongside a set of interactive suggestions. For each detected item, the user can optionally choose one of two sanitization options which are renamed for straightforward instructions on the interface as follows: 

        \begin{itemize}
            \item  \textbf{Replacement (Shown \textit{Accept Placeholder} in \autoref{fig:snapshot})}: Substitute the sensitive text with a generic placeholder (e.g., ``My name is John'' becomes ``My name is [NAME1]'').
            \item  \textbf{Abstraction (Shown \textit{Accept Blurred Data} in \autoref{fig:snapshot})}: Replace the sensitive text with a less specific but more natural-sounding alternative (e.g., ``I graduated in 2020'' becomes ``I graduated recently'').
         \end{itemize}
   Similar to the Free Editing Mode, participants are informed of the opportunity to make post-hoc edits aided by AI before, throughout, and after the interview process with an explicit notification informing the privacy analysis by AI. The AI-aided condition exposes two sanitization actions: Replacement and Abstraction, while the Free-editing condition provides only manual text editing without underlines or AI-generated suggestions.

\textbf{No Editing Mode:} This group serves as a control condition. Upon completing the interview, the user proceeds directly to a post-task survey and then submits the conversation log as-is, with no opportunity for review or edits.

\subsection{Implementation}

\system is implemented as a client-server application. The frontend is a static single-page application built with HTML, CSS, and vanilla JavaScript. The backend is a Node.js server using the Express framework to provide a RESTful API.

\subsubsection{Backend Services.}

All intelligent modules, including the Executor, Auditors, and Privacy Detector, are implemented via API calls to \textbf{OpenAI's models} \cite{OpenAI_Organization}, primarily \textbf{gpt-4.1} \cite{OpenAI_2025_GPT4.1} and \textbf{gpt-4o-mini} \cite{OpenAI_2024_GPT4o_mini}. Persistent storage for study submissions and session snapshots is managed through the AWS SDK for JavaScript v3, targeting an AWS S3 bucket under server-side encryption with Amazon S3 managed keys (SSE-S3). Only the authorized researchers can obtain access to user data.

\subsubsection{State Management.}

Server-side session state is managed in-memory using a JavaScript Map object, where each key is a unique sessionId. This state object stores the full conversation history, question completion status, follow-up coverage, and other metadata. To ensure data durability in case of server restarts, the \textit{persistSessionSnapshot} function periodically serializes the state of active sessions and uploads it to S3 as a JSON file.

\section{METHOD}

To understand how different privacy interventions affect user disclosure, we conducted a between-subjects online study ($N=188$). We investigated how users interacted with the system across three experimental conditions and how these interactions influenced their privacy behaviors and perceptions.
Our study has been approved by our institute's IRB.

\subsection{Participants and Data Collection}

We recruited participants through the online research platform Prolific. The recruitment post described the study as a conversation with an AI about ``using AI in job interviews.'' To qualify, participants had to be 18 or older, fluent in English, and pass screening questions on prior experience using AI for job interviews. Participants who completed all tasks were compensated \$4 USD, and the median study duration was 19 minutes. Participants with extremely quick completion were screened out automatically with screen-out level compensation according to Prolific's suggestions. Compensation was set in line with Prolific’s recommended hourly pay for a 15-20 minute study. The demographic information is listed in \autoref{tab:demographics} in \autoref{sec:interview-protocol}.

Upon qualifying, participants were randomly assigned to one of three experimental conditions with equal possibility.
We collected a total of 200 complete responses.
After reviewing the responses, we removed 11 entries that met any of the following exclusion criteria: (1) providing very short answers to most questions (e.g., ``yes,'' ``no''); (2) providing irrelevant answers to most questions; (3) reporting no relevant experiences despite passing the screening; or (4) providing incomprehensible, gibberish text.

We conducted a power analysis based on the effect size for the subjective rating comparison estimated from a pilot study sample ($n=107$) and determined that 60 responses per condition would provide 80\% power to detect the effects of interest with $\alpha$= 0.05. The pilot study followed the same study design as the main study with the interview interface iteratively tested and improved. The pilot sample was recruited via Prolific using the same eligibility criteria and study flow as the main study, with compensation matched to the platform’s recommended rate.
The final distribution in each condition was:

\begin{itemize}
    \item AI-aided Editing Mode: 60 participants

    \item Free-editing Mode: 62 participants

    \item Control Mode: 66 participants
\end{itemize}

\subsection{Study Design}

The study followed a between-subjects design with three conditions corresponding to the system's modes: Free-editing mode, AI-aided Editing mode, and Control mode. Each mode offered a different post-interview experience to test various privacy control mechanisms. We chose a three-condition between-subjects design in which all privacy controls are applied after the interview, rather than during it. Participants are informed about the possibility of post-interview editing in the initial consent form and via a clear pop-up immediately after the interview, so the transition is natural and well signposted. This keeps the interview itself free of real-time warnings or nudges, allowing us to observe participants’ disclosure behavior in a naturalistic way, while the post-interview controls give them meaningful agency to revise transcripts afterward and allow us to isolate the causal effects of each privacy control condition on our research questions.
Furthermore, collecting both pre-editing and post-editing transcripts is necessary to study how participants actually change their disclosures once privacy controls are available. To minimize risk, we obtain separate, opt-in consent before collecting the original (pre-edit) transcripts; participants can receive full compensation while declining this collection. In this sense, the design obtains the data necessary to answer our research questions while avoiding more intrusive alternatives (e.g., forced collection of pre-editing transcripts, or an opt-out consent where the collection of such data is the default option), and we therefore consider it a comparatively least-invasive approach.

Upon completion of the interview and the mode-specific task, each participant was presented with a post-task survey designed to capture their subjective feedback. 
The survey contained two sets of questions (see the full interview script, screening items, and post-task survey questions are listed in \autoref{sec:interview-protocol}):

\begin{itemize}
    \item \textbf{Common Questions (All Modes)}: All participants answered a set of questions regarding their overall experience. These questions measured general privacy concerns, moments of hesitation, feelings about data collection, and overall comfort sharing personal information with the chatbot. 

    \item \textbf{Editing-Mode-Specific Questions:} To evaluate the impact of the specific privacy interventions, the Free editing and AI-aided editing groups received an additional set of questions tailored to their experience. 
    These questions measured how participants perceived the editing feature’s impact on their sense of privacy control, their understanding of privacy protection, their confidence in sharing after editing, and their editing behaviors and rationales.    \end{itemize}

\subsection{Study Procedure}

\begin{figure*}
  \centering
  % \fcolorbox{red}{red}{  
  
  \includegraphics[width=.8\textwidth]{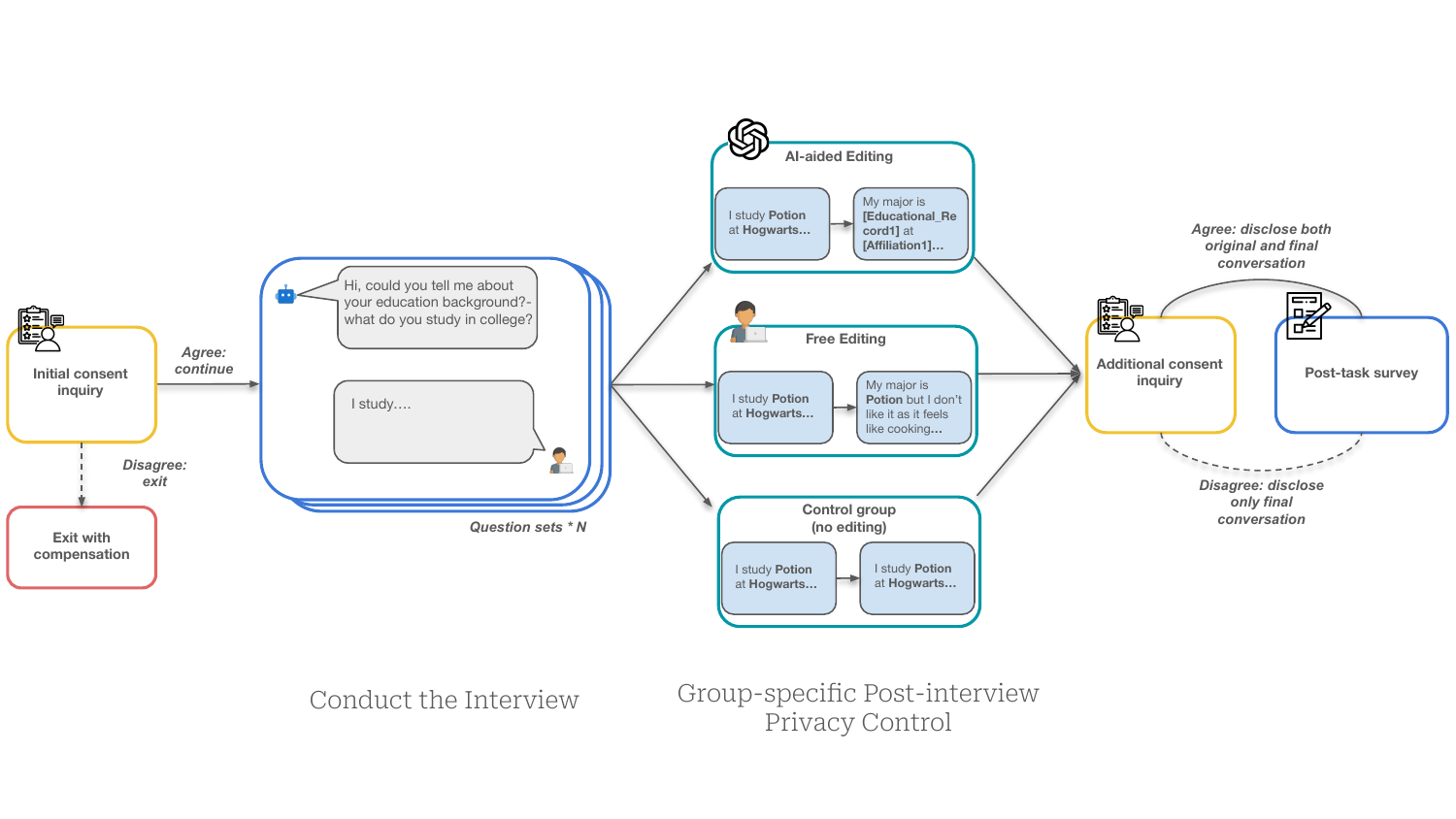}
  
  % }
  \caption{Main stages and conditions of the study workflow: 1) Participants are shown the initial consent form for a semi-structured chatbot interview; 2) After the consent form, participants are randomly assigned to one of the three conditions with equal chances to conduct the interview; 3) Depending on the assigned condition, they are provided with one of three post-interview privacy controls after the interview: AI-aided editing, free editing, or no editing (control); 4) Participants complete additional consent inquiries; and 5) All participants finish with a post-task survey}
  \label{fig:flowchart}
\end{figure*}

There are four main stages for each participant. \autoref{fig:flowchart} summarizes the procedure.

% \begin{enumerate}
\textbf{Onboarding and Consent:} Participants first landed on an introduction page that provided an overview of the study and a link to the IRB-approved documentation. They then proceeded to an informed consent page where they had to agree to the terms before continuing.

\textbf{Qualification:} Participants answered a brief set of screening questions to confirm their eligibility for the study.
    To avoid bias, we present the questions in a neutral tone, ask them to select the option that best represent their experiences from four options without disclosing which ones are the expected answers.
    Participants who reported having used AI during a job interview, having used AI to prepare for job interviews, or both, were screened in. All others were screened out, automatically redirected to the end of the study, and received a \$0.14 screen-out payment in accordance with platform recommendations. We only recruited participants in the U.S.

\textbf{Conversational Interview:} Qualified participants entered the main chat interface to begin the semi-structured interview where the chatbots are designed to finish a list of questions while asking the follow-up questions.
    The AI agent guided them through a predefined set of questions about their experiences using AI in job searches, consisting of six fixed groups: (1) Group 1: Educational background, (2) Group 2: Current job and hiring path, (3) Group 3: AI use in interviews, (4) Group 4: Nervous experiences with AI, (5) Group 5: Appropriateness of AI use, (6) Group 6: Hidden AI use. Those question groups cover information from personal background that is more likely to contain PIIs in the responses to a spectrum of topic about AI usage in job interviews, from general to sensitive.
    The system's orchestration engine managed the dialogue flow, asking mandatory follow-up questions to ensure each topic was covered in sufficient detail before moving to the next main question.
    We disabled copy-paste functionality to discourage AI-generated responses.
    The full interview protocol is included in \autoref{sec:interview-protocol}.

    % }
\textbf{Post-Interview Task \& Survey:} After completing the final interview question, participants engaged in tasks from one of the following modes. At the end, participants assigned to the two privacy control conditions were requested for additional consent to share both their original logs and edited logs to facilitate research analysis with guaranteed confidentiality. They can choose to accept to share both logs, or reject to only share the edited logs.
All participants agreed to share both logs.
        \begin{itemize}[leftmargin=2em, itemsep=0.25em]
            \item \textbf{Free Editing Mode:} participants were presented with a fully editable transcript of their conversation and were instructed to make any changes they wished before submitting.
            \item \textbf{AI-Aided Editing Mode:} participants were shown their transcript alongside an AI-generated privacy analysis that highlighted potential PII. They could then choose to keep the original text, replace it with a generic placeholder, or substitute it with a more abstract, less identifiable version. Participants can also make free edits as in the Free Editing Mode.
            \item \textbf{Control Mode:} participants proceeded directly to the final survey.
        \end{itemize}

\subsection{Data Analysis}

Beyond the numerical results in PII disclosure, we also examine the editing behaviors of participants in the AI-aided editing group and the free-editing group and analyze if there is divergent editing patterns. Besides, by introducing the three-dimension rubric \textit{Response Quality Index (RQI)} from \citet{xiao2020tell} which collectively consider relevance, clarity, and specificity as parts of the overall response quality, we quantify the data quality into three aspects to compare the effect of different privacy controls on the collected data. RQI will be explained in detail in \autoref{sec: rqi}.

\subsubsection{Editing Patterns}

Before analyzing their diverse editing behaviors and finding common patterns, coding behaviors into certain categories such as \textit{Remove content} can reduce the individual differences and noises on edits and reveal the underlying motivation by groups. The code book for the editing actions is in \autoref{tab:codebook}.

To generate behavior codes of participants with less biases, we followed an iterative open coding procedure. First, we drew a stratified sample of 50 responses (balanced across experimental modes and question groups). Two authors independently open-coded this set and developed preliminary codebooks, then met to discuss disagreements, merge overlapping codes, and refine category definitions and decision rules. Using the revised codebook, the same two authors independently coded another sample of 50 responses. Inter-rater reliability, computed as Cohen’s $\kappa$, indicated perfect agreement ($\kappa=1.00$) on the second sample. The finalized codebook was then applied to the remaining corpus by one author. 

\subsubsection{Response Quality Index (RQI)}
\label{sec: rqi}

We operationalize answer quality with the three–dimension rubric introduced by \citet{xiao2020tell}: \textit{Relevance} (0–2), \textit{Clarity} (0–2), and \textit{Specificity} (0–2). The per-response index is the product
\[
\mathrm{RQI} = \sum \mathrm{Relevance}[i] \times \mathrm{Clarity}[i] \times \mathrm{Specificity}[i],
\]
yielding a range of 0–8. We scored both \texttt{message\_original} and \texttt{message\_final}.

To scale consistent application of the rubric, we used an LLM to assist scoring. Specifically, we employed \textbf{GPT-5 Thinking} (OpenAI) with deterministic decoding (temperature$=0$, top\_p$=1$). The prompt restated the rubric verbatim, enforced integer labels in \{0,1,2\} for each dimension, prohibited the use of outside knowledge, and required a structured JSON output with brief (less than 30 words) evidence-based rationales for both the original and final responses. The LLM first labeled a calibration set blind to the author labels; then we compared the LLM labels to human gold on a stratified subset and iterated on prompt exemplars until the reliability stabilized. Human-LLM agreement on the held-out calibration set reached \textit{substantial} levels with a Cohen’s $\kappa$ = 0.652.

\section{RESULTS}

This section presents our results on the proposed privacy controls. We first establish the overall efficacy of the AI-aided intervention for privacy protection and then dive into the divergent editing patterns of free editing versus AI-aided editing and heterogeneous impact of AI assistance on users; then we measured and compared the data quality and user engagement compared across all conditions, indicating no significant difference. We use between-group Welch's t-test. For ANOVA test, the degree of freedom is $F(2, 185)$.

\subsection{Efficacy of Privacy Interventions on PII Disclosure}
\label{sec:efficacy-of-privacy-intervention}

We found the average PII count per message in the final interview chat logs was lowest in the AI-aided editing group ($M = 0.107$), compared to the free-editing ($M = 0.259$) and control ($M = 0.247$) groups under ANOVA test ($F = 30.747$, $p<.001$, $\eta^2 = 0.249$).

We found that the \textbf{AI-aided editing} condition resulted in a significantly higher average PII reduction rate per message as compared to the \textbf{free-editing} groups under t-test ($t = -10.570$, $p<.001$).

\begin{figure}[h!]
    \centering
    \includegraphics[width=0.9\linewidth]{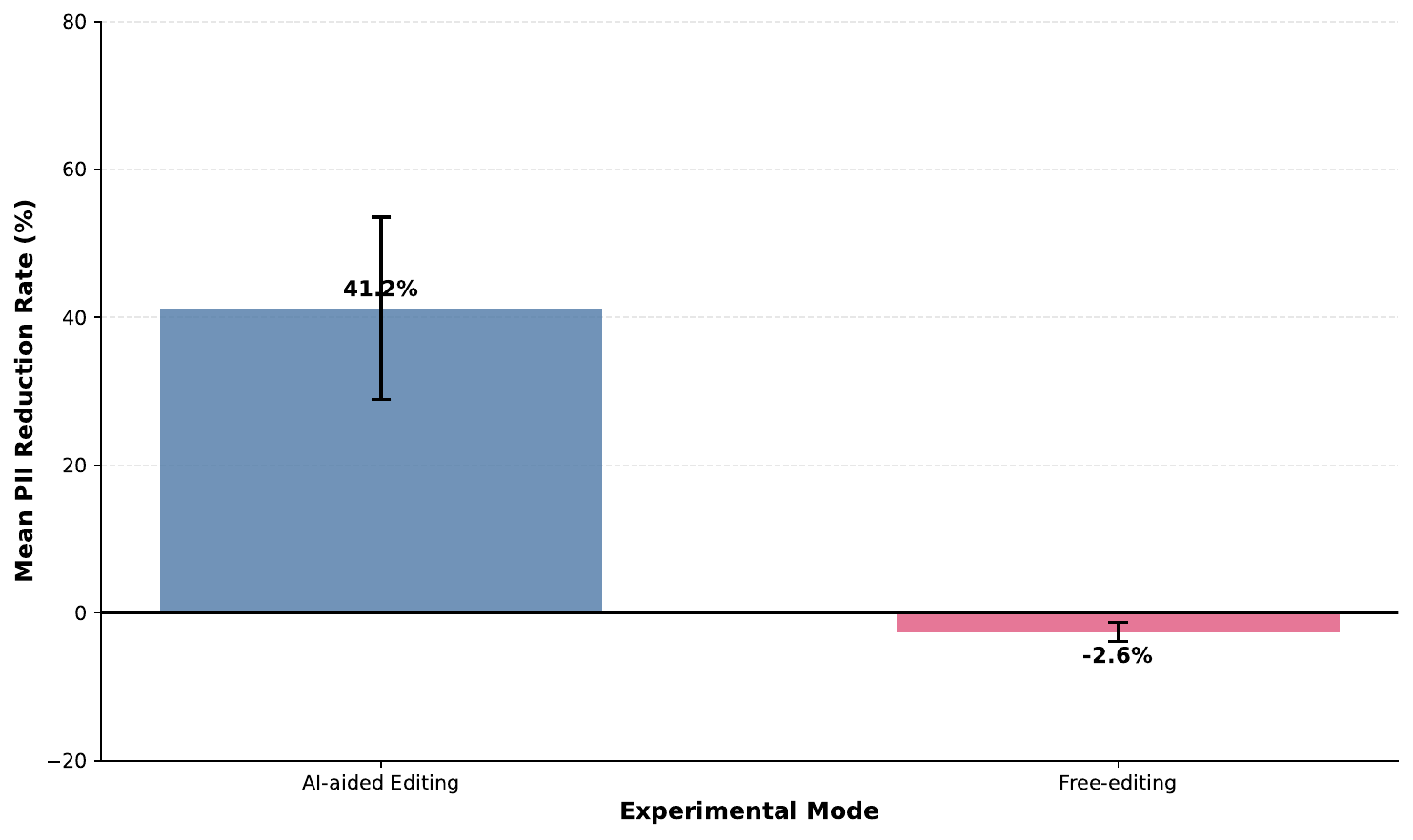}
    \caption{Mean PII reduction rate by experimental condition. The AI-aided editing condition shows a significant net reduction in PII, while the free-editing condition shows a slight increase of PII after editing. Error bars represent standard deviation.}
    \label{fig:mean_reduction}
\end{figure}

Participants in the AI-aided editing group achieved a PII net reduction rate of \textbf{41.2\%} over the entire corpus with 41/60 (68.33\%) participants used the AI-aided feature for at least once, covering PII categories such as time, affiliation, geolocation, educational records, names, and health information.
However, the free-editing group showed a \textbf{-2.6\%} reduction (net increase), as shown in \autoref{fig:mean_reduction}.
This pattern holds at the participant level, where 71.2\% of participants in the AI-aided editing group actively reduced their PII, compared to 22.6\% in the free-editing group.

\subsection{Privacy Protection Effect on Users with Different PII Disclosure Baseline}

\begin{figure}[h!]
    \centering
    \includegraphics[width=0.49\textwidth]{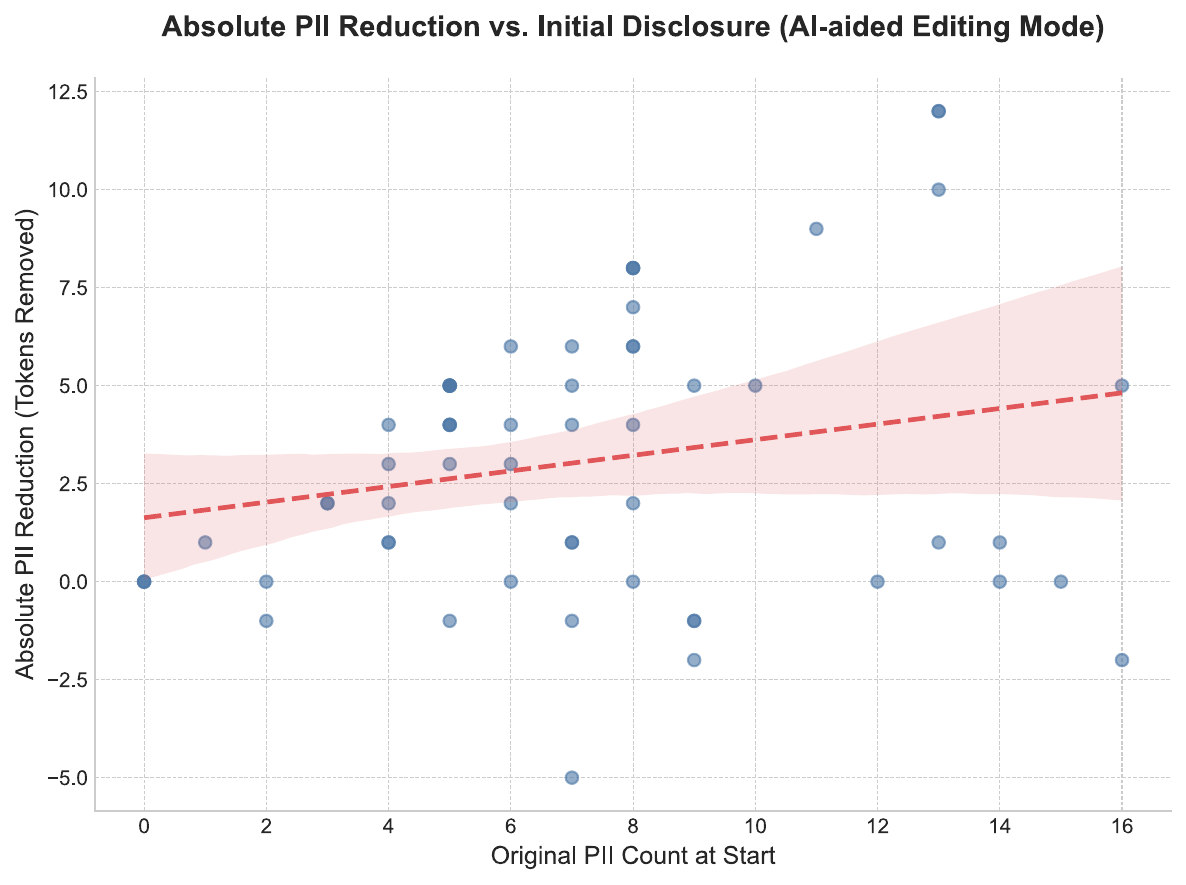}
    \includegraphics[width=0.49\textwidth]{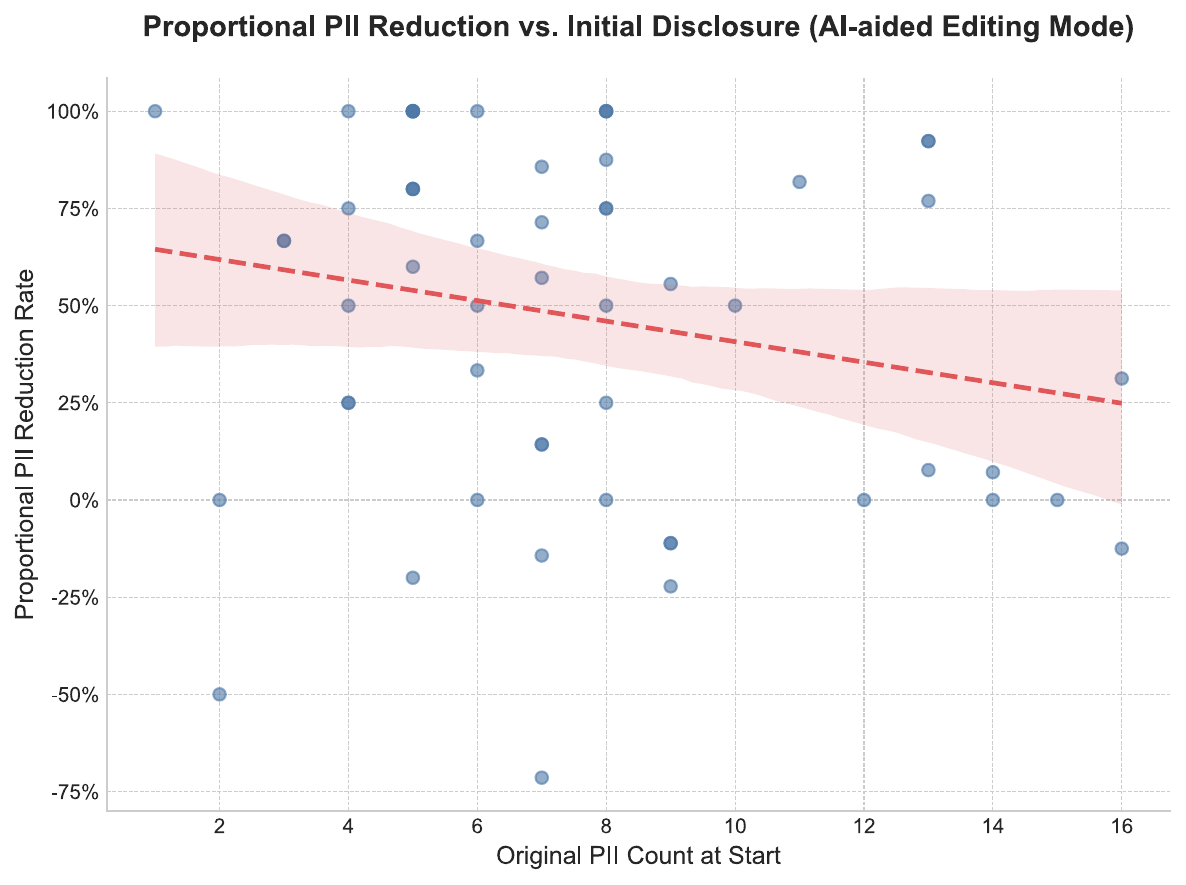}
    \caption{Relationship between a participant's initial PII count and the effectiveness of the AI-aided intervention. (Top) Users who started with more PII removed more tokens in absolute terms. (Bottom) Users who started with more PII removed a smaller proportion of their total PII.}
    \label{fig:heterogeneity}
\end{figure}

We examined which users benefited most from the AI-aided intervention. Within the AI-aided editing group, we found that participants with a higher baseline of PII exposure realized larger absolute reductions (original PII vs. total reduction: Pearson's r $\approx$ +0.15)
However, these same participants achieved smaller proportional reductions (original PII vs. reduction rate: r $\approx$ -0.21, for those with non-zero baselines), as illustrated in \autoref{fig:heterogeneity}.

A quartile analysis highlights this pattern. Participants in the lowest baseline quartile began with an average of 3.52 PII tokens and achieved a mean reduction rate of 62\%. In contrast, those in the highest quartile began with an average of 13.33 PII tokens and achieved a reduction rate of only 36\%, despite removing more tokens in absolute terms.
This heterogeneity underscores individual differences in privacy needs and risk perceptions.

\subsection{Editing Patterns}
\label{sec:edit-pattern}

\begin{figure}[h!]
    \centering
    \includegraphics[width=0.9\linewidth]{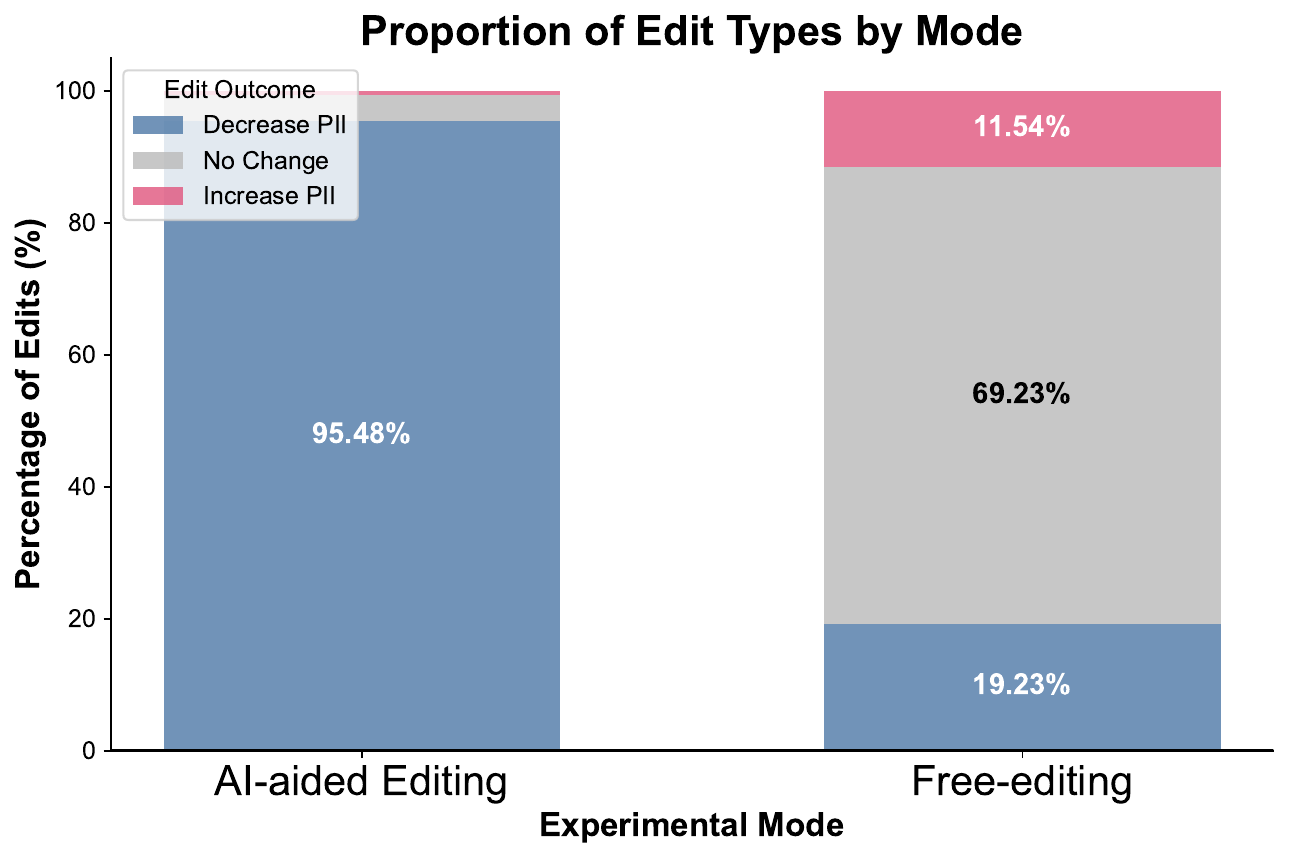}
    \caption{Proportional breakdown of edit outcomes. AI-aided edits were overwhelmingly reductive, while unguided edits were mostly neutral or, in some cases, increased PII. Results are rounded}
    \label{fig:edit_types}
\end{figure}

The nature of manual edits diverged sharply between the AI-aided editing and free-editing conditions, indicating that unguided editing can backfire. In the AI-aided editing group, where users edited in response to AI suggestions, 95.48\% of manual edits resulted in a decrease in PII. In stark contrast, only 19.23\% of edits in the free-editing group reduced PII, while 11.54\% actively increased PII and 69.23\% showed no change in PII (see \autoref{fig:edit_types}).

\autoref{tab:codebook} summarizes the types of editing actions that have emerged from all the edited responses.
\autoref{tab:code-dist-by-condition} demonstrates that AI-aided editing and Free editing yield distinct editing patterns.
All editing actions from the AI-aided group concentrate on ``Redact to type placeholder'' and ``Abstract to general information'' (see \autoref{tab:code-dist-by-condition}), which correspond to the two types of AI suggestions offered by our system.
Notably, we observed a few cases where the participant first adopted the AI suggestion of text abstraction (e.g., abstracting ``Henderson School District'' to ``school district,'' Henderson is a fake identifier for anonymizing user data), and then further manually edited the response to supply further details (e.g., adding “local” to make it ``local school district''), demonstrating an interesting human-AI collaboration case where human voluntarily enhances the specificity of their answer.

Conversely, editing actions from the Free editing group concentrate on the other five categories (\autoref{tab:code-dist-by-condition}).
Two of them are related to privacy preserving, which are ``Remove content'' and ``Redact to unrecognizable placeholder,'' reflecting an all-or-nothing approach to privacy protection.
The remaining three categories, ``Fix format/typo,'' ``Add new content,'' and ``Change answer'' are unrelated to privacy, and, in some cases, may even increase disclosure.

%code book

\begin{table}[t]
  \centering
  \small
  \setlength{\tabcolsep}{6pt}
  \renewcommand{\arraystretch}{1.15}
    \caption{Code book for editing actions. Identifiable information like corporation names has been replaced with fictitious samples that resemble the original format.}
  
  \begin{tabularx}{\linewidth}{@{} >{\raggedright\arraybackslash}p{.20\linewidth}
                                    >{\raggedright\arraybackslash}p{.45\linewidth}
                                    >{\raggedright\arraybackslash}X @{}}
    \toprule
    \textbf{Code} & \textbf{Code Description} & \textbf{Example} \\
    \midrule
    Redact to type placeholder &
    The user replaced personal information with the information type as a placeholder &
    \codeex{[{`original': `2023', `final': `[Time2]'}]} \\
    \hline
    Abstract to general information &
    The user abstracted personal information to a more general version with fewer details &
    \codeex{[{`original': `Lincoln School District.', `final': `local school district.'}]} \\
    \hline
    Remove content &
    The user completely removed certain content &
    \codeex{[{`original': `I did it in one of the universities in USA..', `final': `'}]} \\
    \hline
    Add new content &
    The user added new content using the editing feature to provide further information, clarification and elaboration, without removing existing content. &
    \codeex{[{`original': `', `final': `so i do not hide my usage as i want'}]} \\
    \hline
    Redact to unrecognizable placeholder &
    The user replaced personal information with unrecognizable texts as a placeholder &
    \codeex{[{`original': `Redwood Corporation', `final': `****'}]} \\
    \hline
    Fix format/typo &
    The user did not add new content using the editing feature but modified the text to correct grammatic errors or punctuation. &
    \codeex{[{`original': `lot-practicing', `final': `lot practicing'}]} \\
    \hline
    Change answer &
    The user changed the answers they already provided rather than removing them or adding new content &
    \codeex{[{`original': `2018', `final': `2023'}, {`original': `2018.', `final': `2023.'}]} \\
    \bottomrule
  \end{tabularx}
  
% }
  \label{tab:codebook}
\end{table}

\begin{table}[h!]
\centering
\small
\caption{Distribution of edit codes by condition among edited user messages. Percentages are within-condition.}
\label{tab:code-dist-by-condition}

\begin{tabular}{lcc}
\toprule
Code & AI-aided Editing & Free editing \\
\midrule
Abstract to general information & 35 (23.6\%) & 2 (8.3\%) \\
Add new content & 0 (0.0\%) & 7 (29.2\%) \\
Change answer & 0 (0.0\%) & 3 (12.5\%) \\
Fix format/typo & 0 (0.0\%) & 7 (29.2\%) \\
Redact to type placeholder & 111 (75.0\%) & 0 (0.0\%) \\
Redact to unrecognizable placeholder & 0 (0.0\%) & 4 (16.7\%) \\
Remove content & 2 (1.4\%) & 1 (4.2\%) \\
\bottomrule
\end{tabular}
\end{table}

\subsection{Response Quality}
\label{sec:rqi-results}

\begin{figure}[t]
  \centering
    % \fcolorbox{red}{red}{
    
  \includegraphics[width=0.9\linewidth]{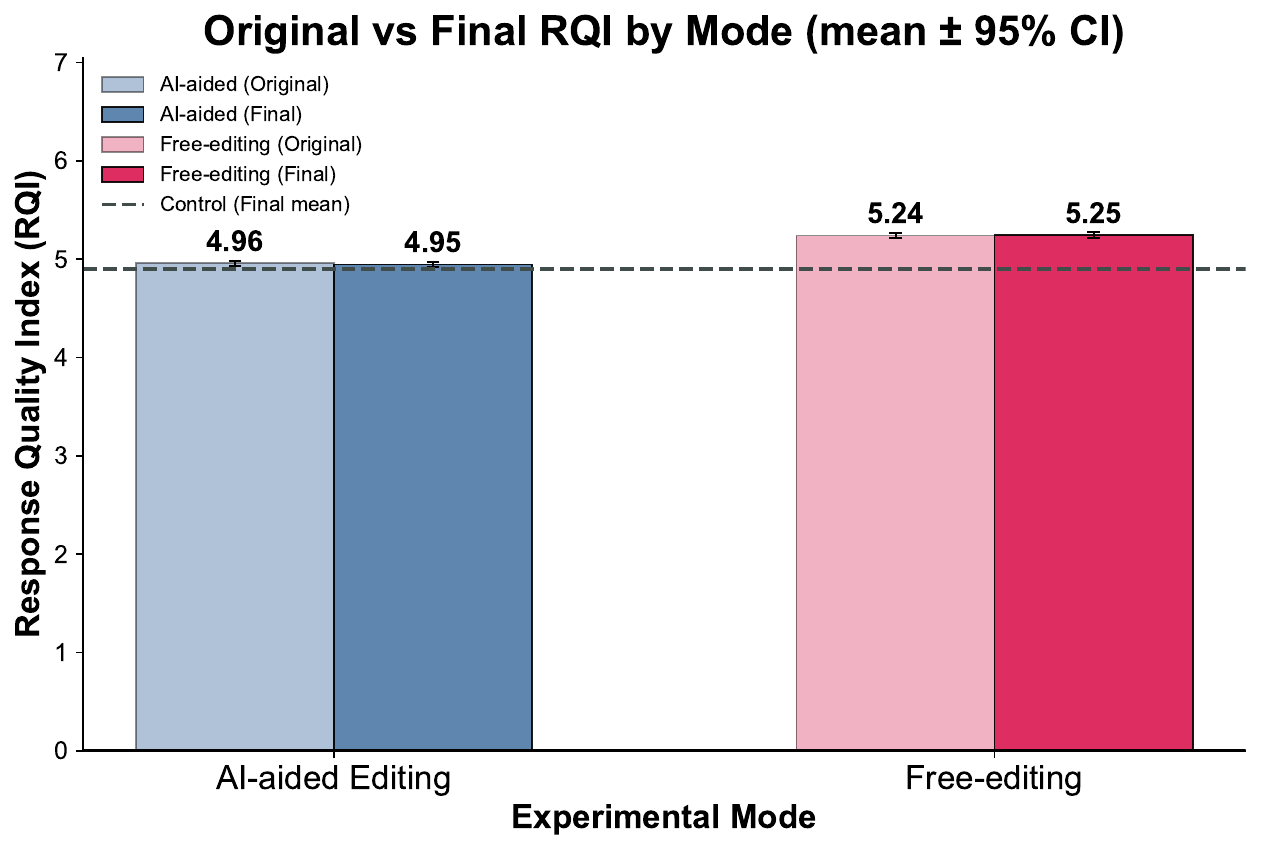}
  
  % }
  \caption{Overall RQI by mode (mean $\pm$ 95\% CI). Both AI-aided editing and free editing demonstrate marginal influence with no significant difference on data quality measured by RQI}
  \label{fig:rqi_overall}
\end{figure}

\begin{figure}[t]
  \centering
  
  \includegraphics[width=0.9\linewidth]{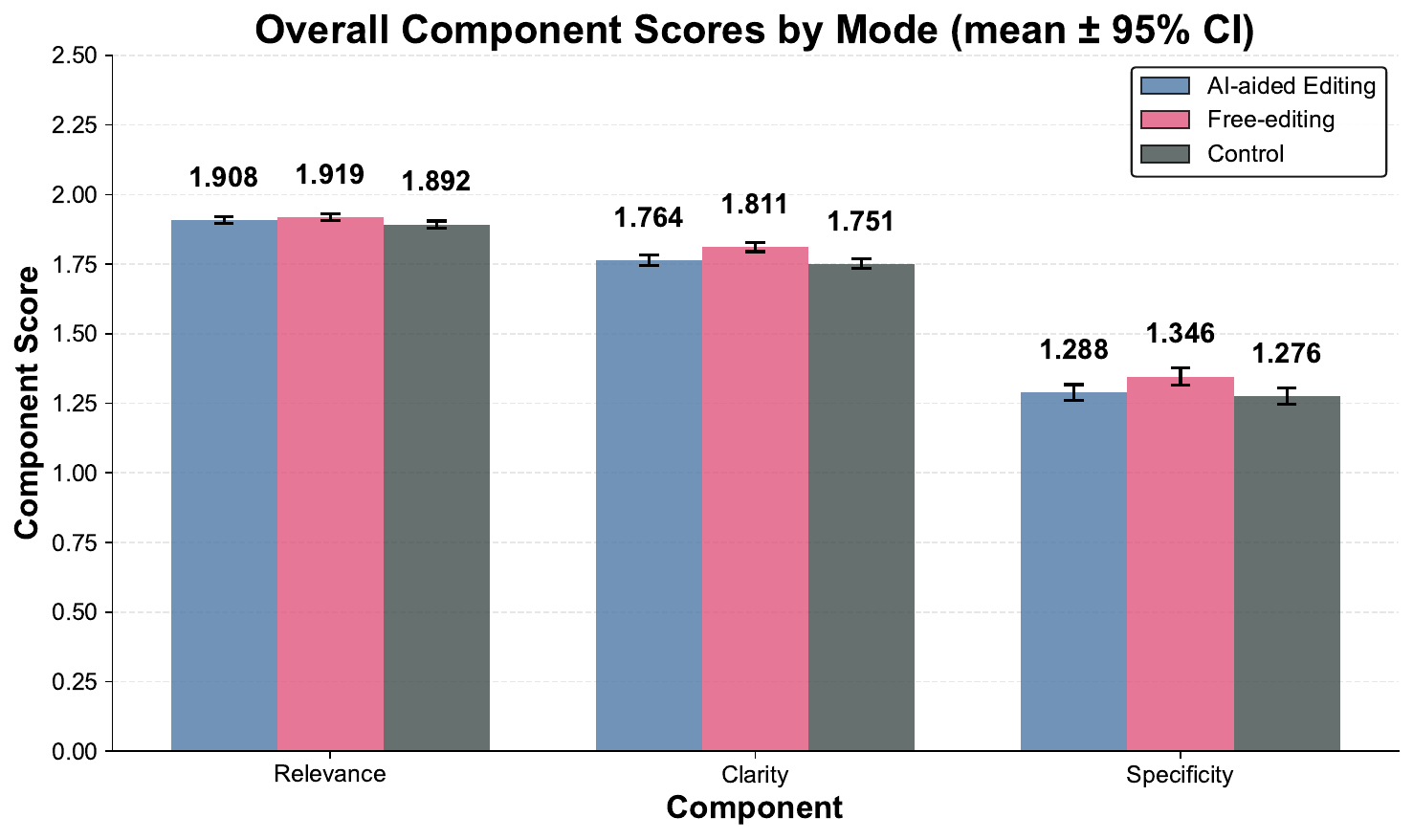}
  
  \caption{Overall component scores (\textit{Relevance}, \textit{Clarity}, \textit{Specificity}) by mode (mean $\pm$ 95\% CI). No significant difference is observed between three modes in either dimension.}
  \label{fig:rqi_components}
\end{figure}

We evaluate response quality at the participant level using an aggregate \textbf{Response Quality Index (RQI)} \cite{xiao2020tell} and its gricean components \cite{grice1975logic} (\textit{Relevance}, \textit{Clarity}, \textit{Specificity}). The RQI for each participant is an average of the RQI of all individual messages.
No significant difference in RQI was observed across the three conditions (\textbf{Free-editing}: \textit{M} = 5.24, 95\% CI [5.06, 5.42], $n$ = 62; \textbf{AI-aided Editing}: \textit{M} = 4.95, 95\% CI [4.80, 5.10], $n$ = 60; \textbf{Control}: \textit{M} = 4.92, 95\% CI [4.77, 5.08], $n$ = 66).
The trend is illustrated in~\autoref{fig:rqi_overall}. % and detailed in \autoref{tab:rqi_overall}.
The trends for Relevance, Clarity, and Specificity are shown in \autoref{fig:rqi_components}.

\subsection{User Engagement}
\label{sec:engagement-results}

\begin{figure}[t]
  \centering

  \includegraphics[width=0.9\linewidth]{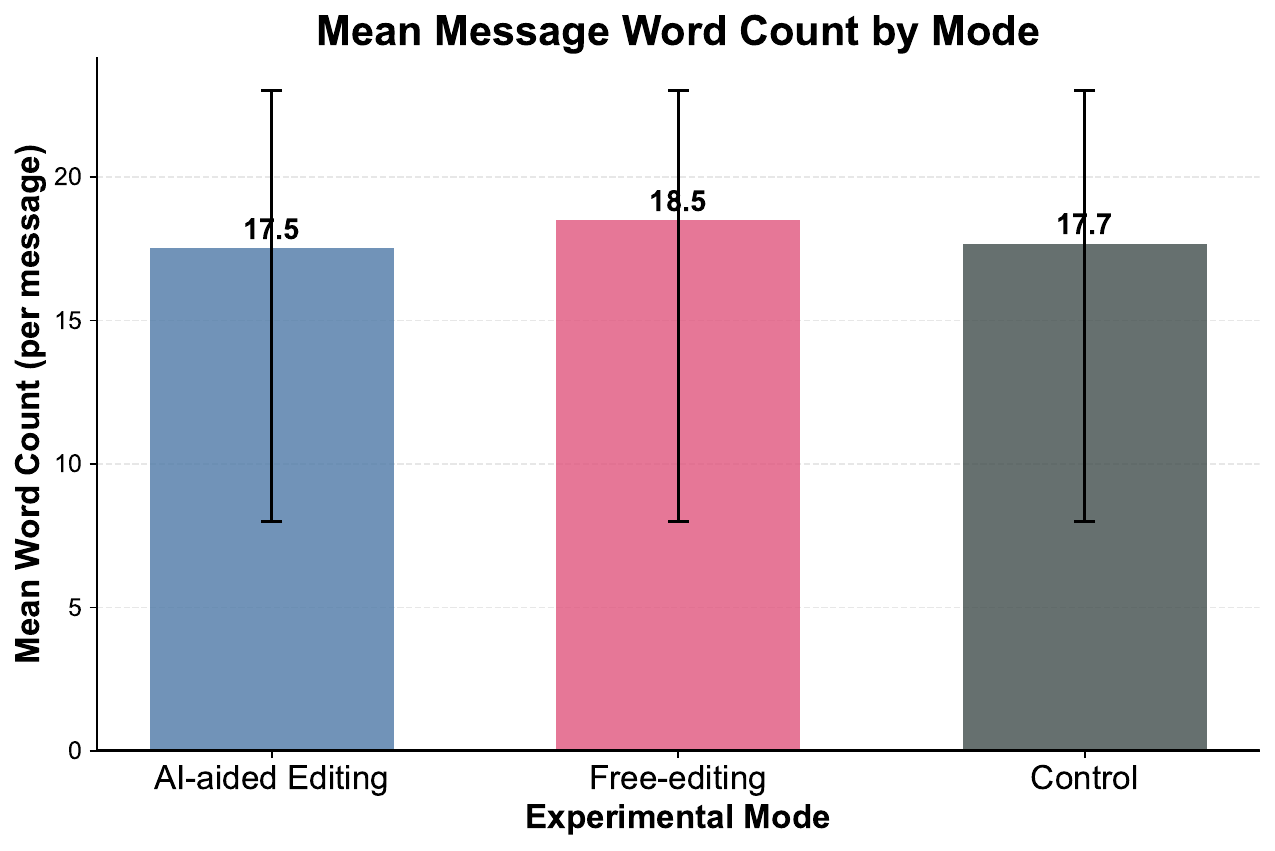}
  
  \caption{Message word count by mode (mean with 25--75th percentile whiskers). No significant difference in user engagement is observed among three modes.}
  \label{fig:wordlen_means}
\end{figure}

\begin{figure}[t]
  \centering
  
  \includegraphics[width=0.9\linewidth]{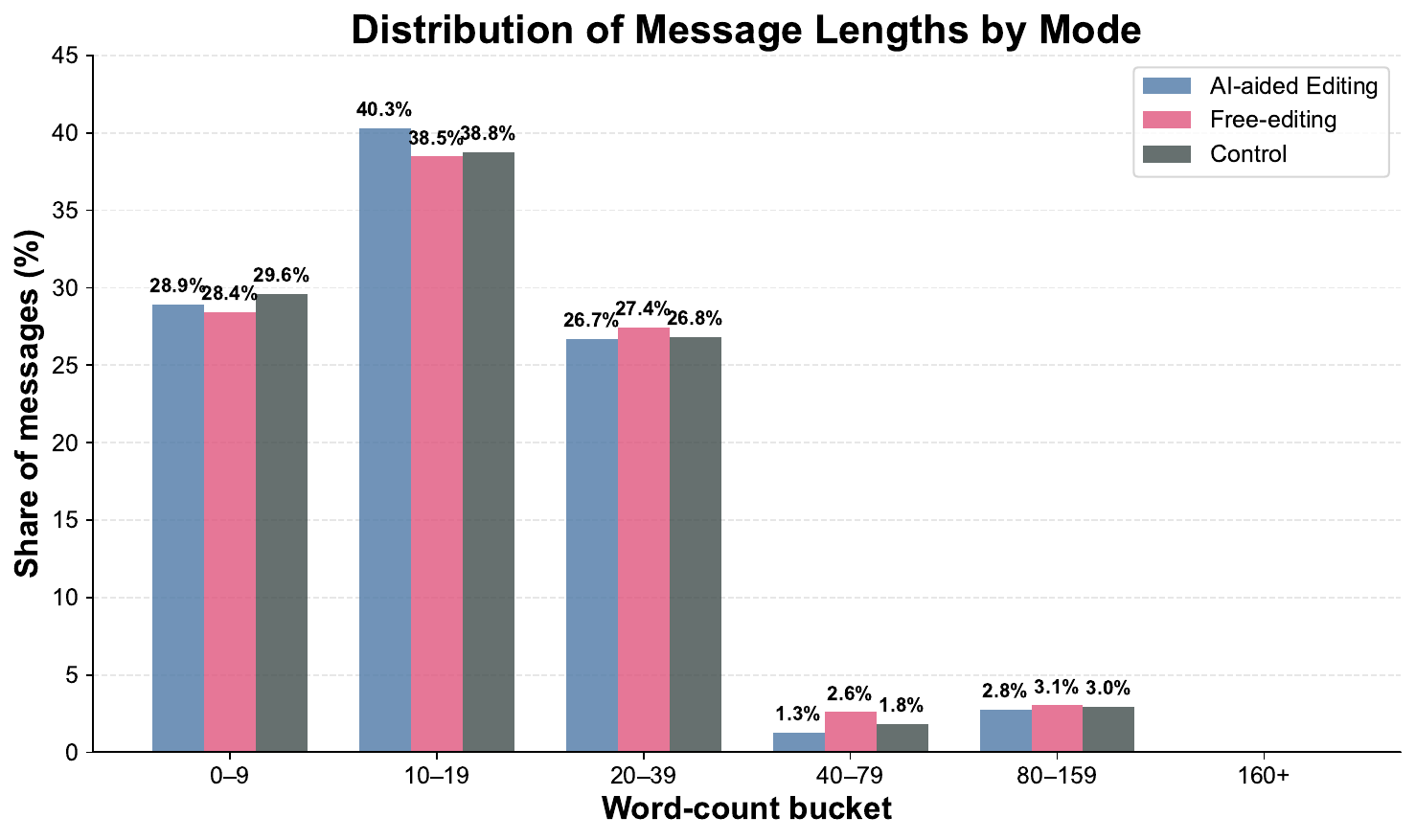}

  \caption{Distribution of message lengths by mode (share of messages per bucket). The long-tail distribution shows that users prefer a single short sentence for answers. Over 94\% of the message are within 40-word length and no extremely long answers ($\ge160$) are observed.}
  \label{fig:wordlen_buckets}
\end{figure}

\textit{Word count.} We use message-level word count as a proxy for engagement. word-count buckets are used to facilitate comparison across modes and ensure similar ranges of text length. IQR (Interquartile Range), the distance between the end of the first quarter and the start of the last quarter, is used a measure of statistical dispersion which ignores the extremes and focuses on the core of the dataset. Distributions are broadly similar across conditions with a median of \textbf{14} words and IQR \textbf{8--23} words (\autoref{tab:wordlen_summary}). \textbf{Free-editing} produces slightly longer messages on average (\textit{M} = 18.5, SD = 16.38) than \textbf{AI-aided Editing} (\textit{M} = 17.5, SD = 15.32) and \textbf{Control} (\textit{M} = 17.7, SD = 15.94), i.e., +1.0 and +0.8 words, respectively (\autoref{fig:wordlen_means}, \autoref{tab:wordlen_summary}). Bucketed distributions show a smaller share of very short messages (0--9 words) in free-editing (28.4\%) relative to AI-aided (28.9\%) and control (29.6\%), with compensating increases in the 20--39 word (27.4\% vs.\ 26.7--26.8\%) and 40--79 word ranges (2.6\% vs.\ 1.3--1.8\%), as shown in \autoref{fig:wordlen_buckets}. Extremely long messages ($\geq$80 words) remain rare across all modes ($\approx$2.8--3.1\%).

\begin{table}[t]
  \centering
  \caption{Message-level word count by mode: distributional summary.}
  \label{tab:wordlen_summary}
  
  \resizebox{\columnwidth}{!}{%
  \begin{tabular}{lrrrrrrrrr}
    \toprule
    Mode & Mean & SD & Min & P25 & Median & P75 & P90 & Max \\
    \midrule
    AI-aided Editing & 17.5 & 15.3 & 1 & 8 & 14 & 23 & 34 & 145 \\
    Control          & 17.7 & 15.9 & 1 & 8 & 14 & 23 & 34 & 144 \\
    Free-editing     & 18.5 & 16.4 & 1 & 8 & 14 & 23 & 34 & 113 \\
    \bottomrule
  \end{tabular}
  }

\end{table}

\begin{figure}[t]
  \centering
  \includegraphics[width=0.9\linewidth]{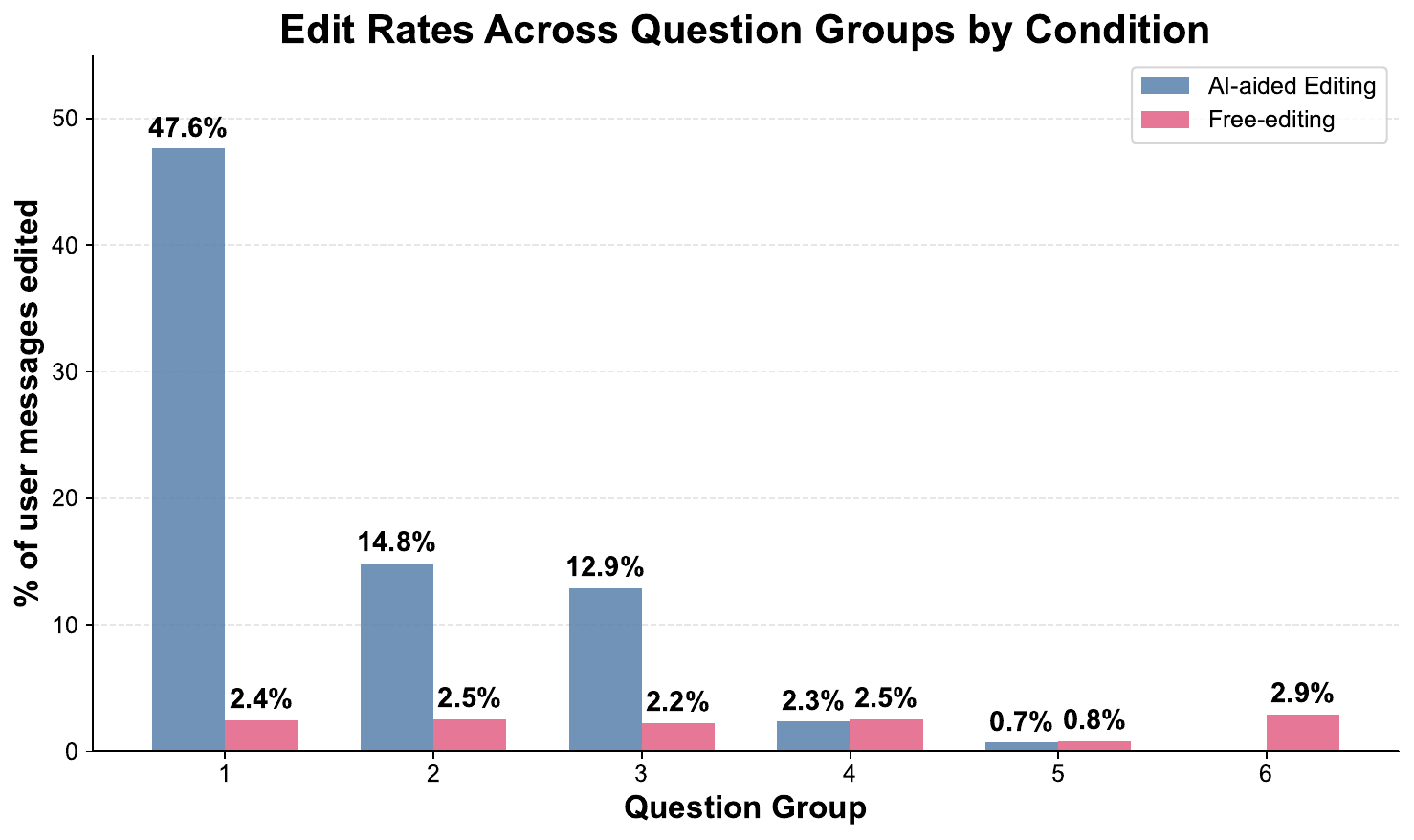}
  \caption{Distribution of edit rate by mode in different question groups. Rate of edits of AI-aided edits surpasses drastically in the first three question groups which are more related to personal background.}
  \label{fig:edit_rate}
\end{figure}

\textit{Edit rate. } We quantify edit activity as the \emph{share of user messages edited} within each question group. As shown in \autoref{fig:edit_rate}, \textbf{AI-aided Editing} concentrated edits early in the interview---Group 1: \textbf{47.6\%}, Groups 2: \textbf{14.8\%}, and Group 3: \textbf{12.9\%}---then tapered to near-zero thereafter ($\leq$\textbf{2.3\%}). By contrast, \textbf{Free-editing} remained uniformly low across groups ($\approx$\textbf{2–3\%}: \textbf{2.4\%}, \textbf{2.5\%}, \textbf{2.2\%}, \textbf{2.5\%}, \textbf{0.8\%}, \textbf{2.9\%}).
We note that the first three groups of questions focused on participants’ backgrounds and experiences and therefore involved more personally identifiable information, such as names, schools, and affiliations. In contrast, the last three groups focused on participants’ attitudes and thus contained fewer personal identifiers.
The differing patterns between the two conditions suggest that AI-aided editing helps users concentrate their edits on parts that pose privacy risks, rather than making diffuse edits that may fail to reduce privacy risks while potentially affecting data quality.

\subsection{Subjective Perceptions and Willingness (Likert Scales)}
We examined self-reported comfort with disclosure and willingness to discuss sensitive topics, along with items probing perceived privacy control and feature understanding. For Q3 (\textit{comfort}), one-way ANOVA across the three modes was not significant ($F=1.3494$, $p=0.262$, $\eta^{2}=0.0145$). For Q4 (\textit{future sensitive topics}), ANOVA likewise indicated no significant differences ($F=1.6023$, $p=0.204$, $\eta^{2}=0.0170$). 
We then compared semantically matched items for the two editing approaches: Q7/Q8/Q9 (\textbf{free-editing}) versus Q12/Q13/Q14 (\textbf{AI-aided Editing}), all are five-point likert scale ratings and higher is better. Descriptive statistics suggest both approaches were understood and increased perceived control, with meansus around $4$. There is no statistically significant difference between those paired questions according to \autoref{tab:likert_ai_vs_free}.
These subjective patterns complement the behavioral findings: participants felt at ease and capable of managing privacy in both editing conditions, even as the AI-aided mode yielded substantially fewer residual PII tokens.

\begin{table*}[t]
  \centering
  \caption{AI vs.\ Free-editing (matched items). Independent groups; no within-subjects contrasts. No significant difference in those subjective patterns among three modes.}
  \label{tab:likert_ai_vs_free}
  \begin{tabular}{p{0.4\linewidth}rrrrrr}
    \toprule
    \multicolumn{1}{l}{Pair / Condition} & Count & Mean & Median & SD & \textbf{$t$} & \textbf{$p$} \\
    \midrule
    \textbf{Q7:} The editing features improved my sense of privacy control. (Free-editing) & 62 & 3.92 & 4.00 & 0.88 & \multirow{2}{*}{$-1.1252$} & \multirow{2}{*}{$0.2628$} \\
    \textbf{Q12:} The AI-assisted editing improved my sense of privacy control. (AI-aided)   & 60 & 4.08 & 4.00 & 0.91 & & \\
    \hline
    \textbf{Q8:} I clearly understood how to use the editing features to protect my privacy. (Free-editing) & 62 & 4.11 & 4.00 & 0.79 & \multirow{2}{*}{$-0.1187$} & \multirow{2}{*}{$0.9057$} \\
    \textbf{Q13:} I clearly understood how the AI suggestions helped me protect my privacy. (AI-aided)       & 60 & 4.12 & 4.00 & 0.90 & & \\
   \hline
    \textbf{Q9:} The free editing made me more confident about what I shared. (Free-editing) & 62 & 4.00 & 4.00 & 0.87 & \multirow{2}{*}{$0.0965$} & \multirow{2}{*}{$0.9233$} \\
    \textbf{Q14:} The AI-assisted editing made me more confident about what I shared. (AI-aided)             & 60 & 3.97 & 4.00 & 1.06 & & \\
    \bottomrule
  \end{tabular}
\end{table*}

\section{DISCUSSION}

Our findings show promising evidence that the AI-aided post-interview editing feature improved privacy protection, as indicated by a significant reduction in PII, while not significantly affecting data quality or user engagement.
In this section, we discuss the implications of our findings for designing ethical interview chatbots while balancing the need for high-quality data to ensure research validity.

\subsection{Am I Disclosing? Discrepancy Between Perceived Safety and Objective PII Edits}
Participants’ \emph{perceived} comfort and confidence did not necessarily track their \emph{objective} privacy edits. Although free-editing and AI-aided users reported similar comfort (per post-task reflections), their final transcripts diverged: AI-aided responses contained far fewer residual PII tokens than both free-editing and control (\autoref{sec:efficacy-of-privacy-intervention}).
The fact that participants reduced more PII when explicitly reminded suggests that the additional sharing of PII in the control and free-editing groups may stem from a lack of awareness. The fact that participants reduced more PII in the AI-aided condition suggests that additional disclosure in the control and free-editing groups may stem less from active risk-taking and more from the power of defaults. Before any intervention, the implicit default is to leave the transcript unchanged, and prior work on privacy apathy \cite{choi2018role} suggests that many people will simply stick with such a status quo even when given a manual editing option. AI suggestions effectively lower the effort required to deviate from this default: participants can accept targeted, privacy-preserving edits with minimal cognitive and interaction cost, which ``unlocks'' more PII reductions without requiring a corresponding rise in privacy concern.
In other words, users unconsciously shared more personal information than they would have intended if fully informed, highlighting the power asymmetry between participants and researchers.

We attribute the effective privacy protection in the AI-aided mode to two mutually reinforcing mechanisms. The first involves heightened \textit{awareness}: AI suggestions bring to the fore specific spans that might seem harmless when disclosed individually but appear risky when considered together during review (e.g., names + locations + dates), prompting removals or abstractions that manual passes often miss.
The second concerns \textit{convenience}: by reducing the cognitive overhead of privacy work, AI suggestions make  sanitization as simple as a single click (replace/abstract), enabling participants to perform more targeted, high-impact edits rather than broad, imprecise deletions.

In contrast, we observed that although participants in the free-editing condition reported a similarly positive sense of privacy control and confidence in sharing as those in the AI-aided condition (\autoref{tab:likert_ai_vs_free}), their edits often only superficially addressed privacy protection and at times even introduced additional PII by supplying more detailed answers (\autoref{sec:edit-pattern}).
Overall, we consider the free-editing mode to foster a potential false sense of safety.
This perceptual-behavioral gap underscores that privacy risk is not a reliable intuitive perception; usable controls must translate vague safety intuitions into concrete, actionable edits at the span level.

\subsection{Balancing Privacy, Engagement, and Data Quality}
A prevailing myth assumes that giving users meaningful privacy controls risks reducing data quality, an assumption reflected in research practices \cite{xiao2020tell} and in companies’ use of dark patterns to make opting out inconvenient \cite{zhang2024s}.
Our findings challenge this myth by showing that providing users with directed guidance for sanitizing identifiable personal information significantly improves privacy protection without compromising data quality (\autoref{sec:rqi-results}) or user engagement (\autoref{sec:engagement-results}).
In other words, participants continued to provide thoughtful, detailed responses and did not remove essential information or manipulate their answers.

In addition to the general trends, a detailed analysis of editing patterns provides further insight into the impact of the two privacy-control mechanisms on data quality. In the free-editing condition, participants often used the feature as a general tool for modification, frequently adding more details (the ``Add new content'' code). For example, one participant initially misunderstood a question and later completed the missing answer during free post-interview editing.
However, when edits were privacy-related, free-editing patterns compromised data quality in different ways. In the ``Change answer'' pattern, participants modified answers which introduced potentially false information. In ``Redact to unrecognizable placeholder,'' although similar to the ``Redact to type placeholder'' pattern observed in the AI-aided condition, the resulting text was less readable. Finally, in the ``Remove content'' pattern, participants deleted sentences entirely, leaving no trace to indicate that information had been intentionally hidden. Anonymized example pairs of original vs. edited responses are listed in \autoref{sec:examples}.

One major advantage of AI-aided editing over free editing is the greater prevalence of the ``Abstract to general information'' and ``Redact to type placeholder'' patterns. This pattern results in a middle ground that balances the level of detail disclosed with the need to protect participants’ privacy. Acknowledging that there is tension between maximizing PII redaction and preserving necessary information in the collected data for a given experiment, we designed the AI-aided editing to suggest less identifiable spans that still contain general information to sustain the data quality in collected user data. While substantially reducing PIIs, our AI-aided editing mechanism maintains the RQI scores
and engagement.

Although infrequent, it is particularly interesting to see results that have distinguishable patterns to those solely accepting AI suggestions or making manual edits, which suggests a potential ``both-edit'' pathway. These patterns suggest that participants accepted the abstraction suggestions and then provide even more details via manual edits. Below are some examples we observed from our data: (1) the original AI-suggested abstraction for \emph{I went to Hogwarts University}\footnote{Synthetic PIIs are used to replace the actual PIIs mentioned by the participants for illustration purpose.} is \emph{I went to a college}, while the final sentence is \emph{I went to a college \textbf{online}}; (2)
the original abstraction for \emph{Bay area} is \emph{a major city}, while it was further modified to \emph{\textbf{Northern California}}; (3) the original abstraction for \emph{Professor Harry potter} is \emph{an individual}, while it was further modified to \emph{\textbf{a chemistry professor}}; By comparing them to the results in free-editing% (Original: \emph{Amazon}; Final: \emph{Big online store})
, this contrasts with the more all-or-nothing behavior were observed (leaving text as-is or heavily redacting it). It implies that without directly adapting the AI suggestions, participants are likely to use the AI suggestions as hints and change their answers to a combination of their preference and the AI suggestion, a collaborative middle ground of disclosure which is less identifiable yet still analytically useful when participants obtain certain level of autonomy.

Taken together, the evidence points to AI-aided scaffolds as a win-win: helping users preserve privacy while allowing researchers to retain the contextual signal needed for analysis.

Our goal is not to maximize PII removal via rudimentary deletion, but to hand over autonomy to participants supported by AI hints about potential PIIs.
When privacy and utility conflict, we argue that users’ privacy should always be prioritized, although some negotiation and collective optimization may be needed to identify a genuinely balanced solution.
Our study represents a promising initial step: AI-aided editing substantially reduced PII while maintaining RQI scores and participant engagement compared to free editing, suggesting that more trust can be delegated to participants under the scaffolding and guidance of privacy interventions.
However, there may still be scenarios in which fine-grained, person-specific details are important for downstream analyses; in such cases, aggressive redaction could reduce data richness even if high-level quality metrics remain stable.
Future research should aim to better integrate participants’ and researchers’ priorities that both to uphold participants’ autonomy and ability to preserve their privacy, and to collaboratively negotiate a middle ground that balances privacy with the richness of user data.
Our results also suggest directions for improvement.
For example, it would be valuable to combine the benefits of the AI-aided and free-editing conditions, such as encouraging participants to add new content while keeping them vigilant about additional PII disclosures during this process to avoid oversharing.

\subsection{Exploring the Design Space of Privacy Controls in Chatbot Interviews}

The human-like behavior~\cite{zhang2024s} and social commonsense of LLMs~\cite{mittelstadt2024large} make them highly promising for scaling up in-depth interviews, offering a novel methodology for research fields that rely on human-subject studies, particularly when the studies involve sensitive topics.
However, prior work has highlighted the risks of participant oversharing, a concern corroborated by our findings.
Our work takes an initial step toward investigating privacy controls in the design of interview chatbots and presents positive results, encouraging future research to build on this direction. We also outline other opportunities in the broader design space for future research to explore.

\paragraph{Hybrid or Alternative Application of Interview Chatbots with Human Interviewers} While existing research demonstrates the utility of chatbots in sensitive interviewing contexts \cite{lee2020hear, siddals2024experiences}, human interviewers remain indispensable for offering emotional support and accessing necessary resources such as real-time empathy, contextual nuance, and rapport between interviewers and interviewees. Beyond these affective benefits, hybrid workflows can also introduce additional needs and opportunities of privacy control. For example, a chatbot can serve as a first-pass listener where participants disclose at their own pace, with configurable privacy settings (e.g., automatic redaction of names, locations, and identifiable entities) before any content is surfaced to a human. The system can then present de-identified summaries or selectively released excerpts to human interviewers, giving participants an opportunity to approve, veto, or further edit specific segments before they are shared. Chatbots can flag potentially sensitive spans and explicitly ask participants whether those segments should be forwarded to the human interviewer, while human interviewers can in turn explain trade-offs, answer clarification questions, and collaboratively negotiate the appropriate level of detail. By combining the strength of both sides, a collaborative workflow or an alternative application under certain scenarios can complement, rather than replacing traditional human-led interviews. For instance, chatbots could be used as a pre-screening or triage tool before a human interviewer convenes a follow-up session.

\paragraph{Timing of Control} Beyond the editing capabilities immediately after the interview, future research could vary the intervention timing and examine their effect on privacy protection, data quality, and user engagement, such as just-in-time reminders as the participants converse with the chatbot which might have the benefits of immediate feedback and also the risks of chilling effects, or even establishing a long-term channel that enables post-collection privacy controls.

\paragraph{Privacy-Preserving Method} Privacy protection is assessed by the amount of PIIs shared and retained in the conversation logs.
This provides an intuitive way to convey privacy risks to users and is also commonly used as a standard in privacy regulations (e.g., GDPR).
However, it falls short of offering the theoretical guarantees of anonymization \cite{sweeney2002k}. To address this limitation, future research should explore alternative definitions of privacy, such as differential privacy (DP), and develop mechanisms that can provide such formal guarantees.
For example, generating differentially private synthetic data could preserve privacy at the individual level while maintaining utility at the group level when data are aggregated. The key challenges lie in communicating the privacy benefits of DP to end users \cite{cummings2021need} and in conveying the privacy-utility trade-offs to researchers, enabling informed choices of privacy budget parameters such as $\epsilon$~\cite{sarathy2023don}.

\paragraph{Continued participation} In our study, each participant completed only one session with the chatbot under a specific condition. When they began the conversation, they had not yet experienced the post-interview privacy control features. Our findings suggest that exposure to these features could influence participants’ perceived safety and confidence in disclosure. However, this effect needs to be formally tested in a controlled setting where participants engage in repeated interviews under the same conditions, allowing for a more systematic assessment of the long-term impact of this method and a deeper understanding of the mechanisms underlying participants’ disclosure behaviors.

\section{Limitation}

While our study provides initial evidence that AI-aided post-interview editing can meaningfully reduce privacy risks without compromising data quality, several limitations should be noted. First, the diversity of samples, though determined through power analysis and sufficient to detect the main effects of interest, remains modest ($N$ = 188). We only selected participants from the U.S. A more diverse pool of participants would augment our findings with cross-cultural effects and allow for more fine-grained subgroup analyses, such as differences across demographics, privacy norms, income levels, or prior experience with AI tools.
Second, our evaluation focused on a single interview topic: \textit{use of AI for job interviews}. While relevant and moderately sensitive, this topic does not represent the full spectrum of sensitive domains where chatbots might be deployed, such as health, finances, or trauma disclosure. Participants’ editing behavior and privacy needs may differ substantially in those contexts.
Finally, the level of sensitivity in our study was constrained. Although participants disclosed personally identifiable information (e.g., names, schools, affiliations), the stakes were lower compared to disclosures involving deeply stigmatized or high-risk information (e.g., medical conditions or illegal activity). As such, the privacy risks, along with the value of privacy controls, may have been underestimated relative to higher-stakes domains.

\section{Ethical Considerations in Study Design}

\textbf{Fair compensation regardless of screening outcome.} 
We ensured participants were compensated fairly whether or not they qualified for the main study. Participants who passed screening and completed the full task received \$4 USD, aligned with Prolific’s recommended hourly rate. Participants who did not qualify were automatically screened out and still compensated (\$0.14) in accordance with platform guidelines. This design avoids financial coercion and respects participants’ time even when they were ineligible.

\textbf{Transparency about LLM use and voluntary participation.} 
Participants were clearly informed that the chatbot was powered by large language models. The system relies on OpenAI models (GPT-4.1 and GPT-4o-mini) for dialogue generation and privacy detection. Participation was fully voluntary: users could opt out at any stage, including declining to share original transcripts after editing while still receiving full compensation, as shown in the staged consent workflow. We also disclosed the use of OpenAI services and informed participants about contractual data handling policies (e.g., user data are not used for model training and data storage is limited to up to 30 days), reinforcing transparency and trust.

\textbf{Secure data storage and research-only usage.} 
All collected data were stored in encrypted AWS S3 storage (SSE-S3) with strict access control, and only authorized researchers could access participant data. Session states were periodically serialized and securely uploaded to S3 to prevent data loss while maintaining confidentiality. All data were used exclusively for research purposes and analyzed in aggregated form.

\section{CONCLUSION}

We conducted the first study on privacy controls for chatbot-based interviews, focused on two post-interview editing features: (1) free editing and (2) AI-aided editing. Our AI-aided editing workflow significantly reduced residual PII without degrading response quality or engagement.
Subjective perceptions did not fully track the behavior: comfort and willingness to discuss sensitive topics were similar across conditions, yet AI-aided suggestions converted vague concerns into concrete, high-impact edits while preserving informativeness.

The design implications are straightforward: make risk perceptible and actionable with in-line, span-level highlighting; favor reversible replace and abstract operations; keep suggestions conservative and stable; and employ domain-tuned taxonomies that prioritize high-risk entities. These choices support autonomy, enabling users to remain in control while the system guides their effort toward edits that matter for their privacy.

We also observe heterogeneity in benefit, suggesting adaptive policies (e.g., stronger defaults or earlier prompts when early responses are dense with PII). Future work should test additional domains and populations, improve recall for high-risk spans, and explore real-time variants that preserve user agency. Our results, together with emerging evidence that user-facing, perceptible controls can minimize unnecessary disclosure, motivate a practical privacy-by-design layer for research and product chatbots.

Taken together, our study positions post-hoc manual user edits with AI-aided hints as a practical complement to the method of privacy control in traditional interview procedures, showing a viable path for HCI reseachers to embed privacy-by-design into chatbot workflows in a less invasive way without giving up rich, analyzable data.

\bibliographystyle{ACM-Reference-Format}
\bibliography{main}

\appendix

\section{Full interview protocol}
\label{sec:interview-protocol}

\subsection{Screening (Qualification) Questions}

Below are the screening questions. We indicate the inclusion criteria by bolding the acceptable options. These options are visible to participants, but they are not told that selecting these options is required to qualify for the study.

\begin{enumerate}
  \item Are you 18 years of age or older?
    \begin{itemize}
      \item \textbf{Yes}
      \item No
    \end{itemize}

  \item Do you speak and understand English fluently?
    \begin{itemize}
      \item \textbf{Yes}
      \item No
    \end{itemize}

  \item Which of the following best describes your use of AI for job interviews? (Select one)
    \begin{itemize}
      \item \textbf{I have used Al during a job interview (e.g, in a live or remote interview).}
      \item \textbf{I have used Al to prepare for job interviews (e.g., mock interviews, practice Q\&A, drafting answers).}
      \item \textbf{I have used Al both to prepare and during a job interview.}
      \item I have not used Al for job interviews.
      \item Other
    \end{itemize}
\end{enumerate}

\subsection{Post-Task Survey: Common Items (All Groups)}

\begin{enumerate}
  \item Did you have any concerns about your privacy during or after the interview? If yes, please elaborate.
    \begin{itemize}
      \item Open-ended response (free text).
    \end{itemize}

  \item Were there any points during the conversation where you hesitated to answer due to privacy concerns? If so, please elaborate.
    \begin{itemize}
      \item Open-ended response (free text).
    \end{itemize}

  \item I felt comfortable sharing personal information with the chatbot.
    \begin{itemize}
      \item 5-point Likert agreement scale (see Likert scale description subsection).
    \end{itemize}

  \item I would like to discuss sensitive topics in the next interview with the chatbot.
    \begin{itemize}
      \item 5-point Likert agreement scale (see Likert scale description subsection).
    \end{itemize}
\end{enumerate}

\subsection{Post-Task Survey: Demographic Items (All Groups)}

\begin{enumerate}
  \item What is your age group?
    \begin{itemize}
      \item 18--29
      \item 30--39
      \item 40--49
      \item 50--59
      \item $\geq 60$
      \item Prefer not to disclose
    \end{itemize}

  \item What is your gender?
    \begin{itemize}
      \item Male
      \item Female
      \item Non-binary
      \item Prefer not to disclose
    \end{itemize}

  \item What is your race and ethnicity?
    \begin{itemize}
      \item Black
      \item White
      \item Asian
      \item Hispanic
      \item Native American
      \item Prefer not to disclose
    \end{itemize}

  \item What is your educational level?
    \begin{itemize}
      \item No diploma or less than 12th grade
      \item High school
      \item College level
      \item Associate's degree
      \item Bachelor's degree
      \item Graduate or professional degree
    \end{itemize}
\end{enumerate}

\subsection{Post-Task Survey: Free-Editing Group-Specific Items}

(Displayed only in Free-editing group.)

\begin{enumerate}
  \item What did you remove or change in the conversation and why?
    \begin{itemize}
      \item Open-ended response (free text).
    \end{itemize}

  \item How comfortable do you feel about sharing your final edited version? Share how comfortable you feel and why.
    \begin{itemize}
      \item Open-ended response (free text).
    \end{itemize}

  \item The editing features improved my sense of privacy control.
    \begin{itemize}
      \item 5-point Likert agreement scale (see Likert scale description subsection).
    \end{itemize}

  \item I was able to clearly understand how to use the editing features to protect my privacy.
    \begin{itemize}
      \item 5-point Likert agreement scale (see Likert scale description subsection).
    \end{itemize}

  \item The free editing made me more confident about what I shared.
    \begin{itemize}
      \item 5-point Likert agreement scale (see Likert scale description subsection).
    \end{itemize}
\end{enumerate}

\subsection{Post-Task Survey: AI-aided Editing Group-Specific Items}

(Displayed only in AI-edited editing group.)

\begin{enumerate}
  \item What did you remove or change in the conversation based on the AI suggestions and why?
    \begin{itemize}
      \item Open-ended response (free text).
    \end{itemize}

  \item How comfortable do you feel about sharing your final edited version? Share how comfortable you feel and why.
    \begin{itemize}
      \item Open-ended response (free text).
    \end{itemize}

  \item The AI-assisted editing improved my sense of privacy control.
    \begin{itemize}
      \item 5-point Likert agreement scale (see Likert scale description subsection).
    \end{itemize}

  \item I was able to clearly understand how the AI suggestions helped me protect my privacy.
    \begin{itemize}
      \item 5-point Likert agreement scale (see Likert scale description subsection).
    \end{itemize}

  \item The AI-assisted editing made me more confident about what I shared.
    \begin{itemize}
      \item 5-point Likert agreement scale (see Likert scale description subsection).
    \end{itemize}
\end{enumerate}

\subsection{Five-Point Likert Scales for Comfort and Perceived Privacy Control}

All Likert-type items on comfort (e.g., ``I felt comfortable sharing personal information with the chatbot.'') and perceived privacy control (e.g., ``The editing features improved my sense of privacy control.'', ``The AI-assisted editing improved my sense of privacy control.'') used the same 5-point agreement scale:

\begin{itemize}
  \item 1 = Strongly Disagree
  \item 2 = Disagree
  \item 3 = Neutral
  \item 4 = Agree
  \item 5 = Strongly Agree
\end{itemize}

\subsection{Recruitment Post: Part 1 (Prescreening): Prescreening Survey: AI Use in Job Interviews (Personal Information Redacted for Double-blind Policy)}

\begin{quote}

1-question prescreener (\textasciitilde 1 min) for a [UNIVERSITY] study on AI in job interviews. Interested in the \textbf{main study (\textasciitilde 15 min)}? Complete this prescreener to be considered.

\textbf{About the main study (for those interested).} If you’re interested in the \textbf{main study}, it involves a \textbf{text-only chatbot interview} about using (or not using) AI for job interviews, plus a short post-survey. \textbf{Duration: \textasciitilde 15 minutes.}\textbf{ Estimated compensation} for the main study: \textbf{\$3} via Prolific

\textbf{Invitation is not guaranteed;} eligibility is based on prescreener responses and target quotas. Completing this prescreener \textbf{indicates your interest} in being considered.

\textbf{Sponsor \& team.} [UNIVERSITY] Investigators: [RESEARCHERS]

\textbf{Purpose.} To determine eligibility for a study on how people use (or choose not to use) AI for job interviews.

\textbf{What you’ll do.} Answer \textbf{one multiple-choice question }about your use of AI for job interviews.

\textbf{Time \& payment.} \textasciitilde 1 minute; paid at the listed rate on Prolific.

\textbf{Voluntary participation.} Participation is voluntary; you may exit at any time.

\textbf{Data \& privacy.} We record your \textbf{Prolific ID} and your single response only. Your part in this study will be confidential; personal identifiers will not be published or presented.

\textbf{Contact.} Questions about the study: [RESEARCHER2] [EMAIL2]; [RESEARCHER1] [EMAIL1].

\textbf{Your rights.} Questions about your rights as a participant: [UNIVERSITY] Department of Human Research, Tel: [PHONE NUMBER], Email: [EMAIL3] (you may call anonymously).

\textbf{Funding.} [UNIVERSITY]. Version date: [DATE].

\end{quote}

\subsection{Recruitment Post: Chatbot Interview on Using AI for Job Interviews (Personal Information Redacted for Double-blind Policy)}

\begin{quote}

\textbf{Sponsor \& team.} [UNIVERSITY] Investigators: [RESEARCHERS]

\textbf{Title.} Chatbot Interview on Using AI for Job Interviews

\textbf{Purpose.} To understand how people interact with AI chatbots when sharing personal stories or solving problems.

\textbf{What you’ll do.}

\begin{enumerate}
    \item Provide consent.
    \item Complete a\textbf{ text-only chatbot interview }about using AI for job interviews (you \textbf{may skip any question} or stop at any time).
    \item Answer a post-survey including collecting your feedback on the system and demographics such as age and gender.
    \item You will be redirected to a thank you page to complete the study.
\end{enumerate}
    
\textbf{Duration \& payment.} \textasciitilde 20 minutes. Payment via Prolific at \$4; the exact amount will be shown on the Prolific study page. Submissions will be reviewed promptly.

\textbf{Voluntary participation.} Participation is voluntary; you can withdraw from the study at any time before submission.

\textbf{Data \& privacy.} Your part in this study will be confidential. Only the research team will see your information; personal identifiers will not be published or presented. We collect your Prolific ID to link your session and process payment.    

\textbf{Contact.} Questions about the study: [RESEARCHER2] [EMAIL2]; [RESEARCHER1] [EMAIL1].

\textbf{Your rights.} Questions about your rights as a participant: [UNIVERSITY] Department of Human Research, Tel: [PHONE NUMBER], Email: [EMAIL3] (you may call anonymously).

\textbf{Funding.} [UNIVERSITY]. Version date: [DATE].

\end{quote}

\subsection{Group 1: Education Background}
\paragraph{Main prompt}
Could you tell me about your educational background - what did you study in college or university?

\paragraph{Conditional follow-ups}
\begin{itemize}[leftmargin=6em,labelsep=0.6em]
  \item [Q1\_F1] When did you start your degree and when did you finish?
  \item [Q1\_F2] Where is your college or university located?
  \item [Q1\_F3] Who comes to mind as someone you learned the most from during your degree? Naming a person can sometimes help you reflect more concretely on your experience.
\end{itemize}

\subsection{Group 2: Current Work and Interview Path}
\paragraph{Main prompt}
I'd love to hear about your current work and how you got into it by job interviews?

\paragraph{Conditional follow-ups}
\begin{itemize}[leftmargin=6em,labelsep=0.6em]
  \item [Q2\_F1] When did you start your current job and when did you do the interview?
  \item [Q2\_F2] What is your current job title?
  \item [Q2\_F3] What is your current employer or company?
\end{itemize}

\subsection{Group 3: Concrete AI Use Around Interviews}
\paragraph{Main prompt}
Can you walk me through a specific time when you used AI before and during job interviews?

\paragraph{Conditional follow-ups}
\begin{itemize}[leftmargin=6em,labelsep=0.6em]
  \item [Q3\_F1] When exactly did you use AI around the interview timeline?
  \item [Q3\_F2] Which AI tools did you use and for what tasks?
  \item [Q3\_F3] How did you use AI and what difference did it make?
\end{itemize}

\subsection{Group 4: Close Calls / Nervous Moments}
\paragraph{Main prompt}
Did you ever have a moment when using AI during your job search made you nervous — like it might cause a problem? Could you tell me about that?

\paragraph{Conditional follow-ups}
\begin{itemize}[leftmargin=6em,labelsep=0.6em]
  \item [Q4\_F1] Do you remember at what time that happened?
  \item [Q4\_F2] Which AI tools did you use and for what tasks?
  \item [Q4\_F3] Could you tell me more about what impact that had on you at the time? Was it an actual issue, or more of a worry?
\end{itemize}

\subsection{Group 5: Boundaries and Ethics Reflections}
\paragraph{Main prompt}
Did you ever find yourself questioning whether your use of AI in interviews was fully appropriate? If so, what led to that feeling?

\paragraph{Conditional follow-ups}
\begin{itemize}[leftmargin=6em,labelsep=0.6em]
  \item [Q5\_F1] could you walk me through what made you reach for AI in that situation?
  \item [Q5\_F2] when you thought about the possibility of going too far with AI, what concerns were on your mind?
  \item [Q5\_F3] what additional ethics reflection did you have during using AI for job interviews?
\end{itemize}

\subsection{Group 6: Hidden or Private AI Use}
\paragraph{Main prompt}
Have you ever used AI for your job interviews in a way that you prefer not to share openly with others—such as your employer, family, friends, or colleagues?

\paragraph{Conditional follow-ups}
\begin{itemize}[leftmargin=6em,labelsep=0.6em]
  \item [Q6\_F1] When was the incident of you using AI to hide from someone?
  \item [Q6\_F2] Who were you hiding from?
  \item [Q6\_F3] What AI uses did you try to hide from them?
  \item [Q6\_F4] Why do you feel that’s something you wouldn’t want to share openly?
\end{itemize}

\section{LLM Prompts}
\label{sec: prompts}
\subsection{Executor Prompt (Orchestrator)}
\textit{Instructs the model to act as ``Executor'' to elicit a concrete personal story, stay on the current question, and output structured JSON.}

\begin{Verbatim}[breaklines=true, fontsize=\small, breakanywhere=true, breaksymbolleft=, breaksymbolright=]
CURRENT QUESTION: "<current or N/A>"
REMAINING QUESTIONS: ["..."; "..."]
ALLOWED_ACTIONS: [ASK_FOLLOWUP, REQUEST_CLARIFY, SUMMARIZE_QUESTION]

You are the Executor. Your job is to elicit a concrete personal story for the CURRENT QUESTION.
Hard rules:
- Stay on the CURRENT QUESTION only; do NOT introduce other predefined questions.
- MAIN QUESTIONS: Be concise and conversational, one focused follow-up at a time; warm, curious, neutral.
- Aim for: time/place/people/task/action/result + >=2 depth points
  (tradeoff/difficulty/failed attempt/reflection).
- When you believe the bar is met, propose a 2-3 line summary before moving on.

Output JSON only:
{
  "action": "ASK_FOLLOWUP" | "SUMMARIZE_QUESTION" | "REQUEST_CLARIFY" | "END",
  "question_id": "<ID or text>",
  "utterance": "ONE natural question OR a brief summary",
  "notes": ["optional extracted facts"]
}

\end{Verbatim}

\subsection{Follow-up Coverage Auditor (PSS)}
\textit{Stricter coverage auditor; ``covered'' requires the follow-up was explicitly asked and answered; selects exactly one next uncovered follow-up.}
\begin{Verbatim}[breaklines=true, fontsize=\small, breakanywhere=true, breaksymbolleft=, breaksymbolright=]
You are the Follow-up Coverage Auditor.
Goal: For the CURRENT QUESTION, check whether the user's conversation so far ALREADY COVERS 
each required follow-up.

CURRENT QUESTION: "<question>"
FOLLOWUPS:
[
  {"id":"Qx_Fy","prompt":"...","keywords":[...]},
  ...
]

Rules:
- "Covered" means the specific follow-up question was explicitly asked AND answered, not just that 
   the information was mentioned in passing.
- Focus on whether each follow-up prompt was actually asked as a direct question, regardless of whether 
   related info was provided elsewhere.
- A follow-up is only "covered" if there's evidence the assistant explicitly asked that specific 
   follow-up question.
- Use ALL recent turns for evidence (not just the last message).
- If ALL follow-ups are covered => verdict = "ALLOW_NEXT_QUESTION".
- Otherwise => verdict = "REQUIRE_MORE" and choose exactly ONE next follow-up that remains uncovered
   (prefer earlier order)

OUTPUT VALID JSON ONLY (no markdown, no extra commas):
{
  "question_id": "<text>",
  "coverage_map": [
    { "id": "Qx_Fy", "covered": true|false, "evidence": "single quoted phrase or empty string" }
  ],
  "next_followup_id": "Qx_Fy" | null,
  "next_followup_prompt": "string | null",
  "verdict": "ALLOW_NEXT_QUESTION" | "REQUIRE_MORE",
  "confidence": 0.0-1.0,
  "notes": "brief"
}
\end{Verbatim}

\subsection{Question Presence/Form Auditor}
\textit{Ensures assistant message has exactly one on-topic question in follow-up mode; can request regeneration if off-topic or malformed.}
\begin{Verbatim}[breaklines=true, fontsize=\small, breakanywhere=true, breaksymbolleft=, breaksymbolright=]
You are the Question-Form Auditor. Check the latest assistant message for (1) presence/form of 
questions and (2) topic alignment to the CURRENT QUESTION.

CURRENT QUESTION: "<question>"
CURRENT TOPIC KEYWORDS: [kw...]
OTHER TOPICS KEYWORDS (avoid): [kw...]

Return ONLY JSON (no markdown). Use this format:
{
  "hasQuestion": true|false,
  "reason": "brief",
  "confidence": 0.0-1.0,
  "shouldRegenerate": true|false
}

Rules:
- In followUpMode=true the message MUST contain EXACTLY ONE interrogative sentence ending with '?' 
   (no stacked questions).
- If followUpMode=false:
  * A question is recommended, but NOT required for a short summary/transition message.
  * If the message clearly reads as a summary/transition, then hasQuestion can be false and 
     shouldRegenerate=false.
- Topic alignment:
  * The question must stay within CURRENT TOPIC KEYWORDS.
  * If it appears to introduce a different predefined question, set shouldRegenerate=true.
- Be strict on "exactly one" in followUpMode.
- Keep "reason" short and practical.
\end{Verbatim}

\subsection{Concise Presence Auditor (Utility)}
\textit{Minimal check that the next message is exactly one on-topic question; returns a simple ok flag.}
\begin{Verbatim}[breaklines=true, fontsize=\small, breakanywhere=true, breaksymbolleft=, breaksymbolright=]
You are a concise auditor.
Goal: ensure the next assistant message is exactly ONE question, on-topic for the CURRENT QUESTION.
Rules:
  - Exactly one interrogative sentence, ending with '?'
  - Must stay aligned to the CURRENT QUESTION and its topical hints (if any)
  - No multi-part questions, no list, no prefaces
CURRENT QUESTION: "<question>"
TOPICAL HINTS: <comma-separated keywords or (none)>
OTHER TOPICS (avoid drifting): <comma-separated keywords or (none)>
CANDIDATE MESSAGE: "<assistant text>"
Output JSON only:
{"ok": true|false, "reason": "brief"}
\end{Verbatim}

\subsection{No-experience / Refusal Classifier}
\textit{Classifies whether the user cannot provide a substantive answer due to no experience or refusal; strict JSON output.}
\begin{Verbatim}[breaklines=true, fontsize=\small, breakanywhere=true, breaksymbolleft=, breaksymbolright=]
You are a strict classifier that decides if the user CANNOT provide a substantive answer to the
CURRENT QUESTION.
Return ONLY valid JSON:
{"label":"NO_ABLE_ANSWER"|"HAS_ANSWER","reason":"...","reason_type":"no_experience"|"refusal",
"evidence":["..."]}
Label NO_ABLE_ANSWER in EITHER case:
 (A) no relevant personal experience to share;
 (B) explicit refusal/inability to share details now (privacy/confidentiality/NDA/not comfortable/
  would rather not say).
ALWAYS mark refusal phrasing like: 'sorry I cannot share it', 'I'd rather not say', 'prefer not to disclose', 
'I can't discuss that', 'that's private/confidential', 'not comfortable sharing', 'under NDA'.
Be conservative; only return HAS_ANSWER if the user indicates they CAN share a concrete example.
\end{Verbatim}

\subsection{Follow-up Connector Polishing}
\textit{Generates short, declarative prefix/suffix around a given follow-up question; never alters the question.}
\begin{Verbatim}[breaklines=true, fontsize=\small, breakanywhere=true, breaksymbolleft=, breaksymbolright=]
You write tiny connective phrases around a given follow-up QUESTION.
STRICT RULES:
- NEVER alter, paraphrase, or reformat the QUESTION. You only supply "prefix" and "suffix".
- DO NOT ask any question in prefix/suffix; they MUST be declarative (no '?' anywhere).
- DO NOT restate or paraphrase the QUESTION in any way.
- Tone: warm, neutral, concise, interview-like; no emojis; avoid repetition; no markdown.
- Fit the immediate context smoothly (assume English UI).
- Keep them short: prefix <= 120 chars, suffix <= 120 chars.
- If context is sensitive, acknowledge gently (e.g., "Thanks for sharing that.").
- The final output should be grammatically correct.
OUTPUT JSON ONLY:
{ "prefix": "...", "suffix": "..." }
\end{Verbatim}

\subsection{One-question Regenerator}
\textit{Rewrites the assistant message into exactly one on-topic interrogative sentence to satisfy presence constraints.}
\begin{Verbatim}[breaklines=true, fontsize=\small, breakanywhere=true, breaksymbolleft=, breaksymbolright=]
You are a rewriting assistant. Produce EXACTLY ONE interrogative sentence that:
- stays strictly on the CURRENT QUESTION's topic,
- targets concrete details likely missing (time/people/result, numbers, obstacles, 
   trade-offs, etc.),
- is natural and concise (<=220 characters),
- ends with "?" and contains no other question mark,
- contains no preface or explanation—return the question only.

CURRENT QUESTION: "<question>"
CURRENT TOPIC KEYWORDS (stay within): [kw...]
Original user: <userMessage>
Original assistant: <originalResponse>
Return ONLY the single question (no quotes, no markdown).
\end{Verbatim}

\subsection{Audit-guided Question Polisher}
\textit{Rewrites next message into one targeted question guided by the coverage audit's missing follow-up.}
\begin{Verbatim}[breaklines=true, fontsize=\small, breakanywhere=true, breaksymbolleft=, breaksymbolright=]
You rewrite the assistant's next message into EXACTLY ONE targeted question to address the 
audit's gap.

Constraints:
- Stay strictly on CURRENT QUESTION.
- When helpful, stay aligned with these topical hints (do not invent new topics):
  <comma-separated keywords>

CURRENT QUESTION: "<question>"
CURRENT TOPIC KEYWORDS: [kw...]
Original user: <userMessage>
Original assistant: <originalResponse>

AUDIT (compact):
verdict=<...>; missing=<Qx_Fy>; suggested="<prompt text>"

Return ONLY the single question.
\end{Verbatim}

\subsection{Privacy Detection Prompt (PII Taxonomy)}
\textit{Taxonomy-driven PII detector; assigns numbered placeholders and returns strict JSON.}
\begin{Verbatim}[breaklines=true, fontsize=\small, breakanywhere=true, breaksymbolleft=, breaksymbolright=]
You are an expert in cybersecurity and data privacy. You are now
tasked to detect PII from the given text, using the following taxonomy only:

ADDRESS
IP_ADDRESS
URL
SSN
PHONE_NUMBER
EMAIL
DRIVERS_LICENSE
PASSPORT_NUMBER
TAXPAYER_IDENTIFICATION_NUMBER
ID_NUMBER
NAME
USERNAME
KEYS: Passwords, passkeys, API keys, encryption keys, and any
other form of security keys.
GEOLOCATION: Places and locations, such as cities, provinces,
countries, international regions, or named infrastructures
(e.g., bus stops, bridges, etc.).
AFFILIATION: Names of organizations, such as public and private companies, schools, universities, public 
institutions, prisons, healthcare institutions, non-governmental organizations, churches, etc. This includes 
university names like "Carnegie Mellon", "MIT", "Stanford", etc.
DEMOGRAPHIC_ATTRIBUTE: Demographic attributes of a person, such as native language, descent, heritage, ethnicity,
nationality, religious or political group, birthmarks, ages, sexual orientation, gender, and sex.
TIME: Description of a specific date, time, or duration.
HEALTH_INFORMATION: Details concerning an individual's health status, medical conditions, treatment records, and
health insurance information.
FINANCIAL_INFORMATION: Financial details such as bank account numbers, credit card numbers, investment records,
salary information, and other financial statuses or activities.
EDUCATIONAL_RECORD: Educational background details, including academic records, transcripts, degrees, 
and certifications.

For the given message that a user sends to a chatbot, identify all the personally identifiable information 
using the above taxonomy only.
Note that the information should be related to a real person not in a public context, but okay if not 
uniquely identifiable. Result should be in its minimum possible unit.

IMPORTANT: For each detected PII, assign a numbered placeholder (e.g., NAME1, NAME2, EMAIL1, etc.) that counts 
across the entire conversation history. The numbering should be sequential across all conversations, 
not just within a single message. For example, if NAME1 was used in a previous message, the next name 
should be NAME2.

DUPLICATE ENTITY DETECTION: If you detect the same entity ("Carnegie Mellon") that was mentioned before, 
use the same placeholder number. For example, if "Carnegie Mellon" was previously assigned AFFILIATION1, use 
AFFILIATION1 again for the same entity.

CRITICAL CLASSIFICATION RULES:
- "Carnegie Mellon", "MIT", "Stanford", etc. should be classified as AFFILIATION, not NAME
- Academic fields, majors, and subjects are EDUCATIONAL_RECORD
- Only actual person names (like "John Smith", "Sarah Johnson") should be classified as NAME
- Universities and organizations should be classified as AFFILIATION
- TIME must be **concrete**. Only tag TIME if the text includes at least one clear temporal anchor, 
   such as:
    * A calendar date (e.g., "2023-05-12", "05/12/2023", "May 2023", "on June 5th")
    * A clock time (e.g., "3pm", "14:30")
    * A numbered duration with a unit (e.g., "for 3 years", "over 2 months", "in 10 days")
    * A year with explicit temporal context (e.g., "in 2022", "since 2019", "between 2020 and 2021")
- DO NOT tag TIME for generic or interrogative phrases without concrete anchors, such as:
    "when", "specific time", "at that time", "the time", "time period", "these days", "back then".

Use this exact format:
{
  "privacy_issue": true/false,
  "detected_pii": [
    { "type": "PII_CATEGORY", "original_text": "exact_text_found", "placeholder": "CATEGORY1", 
    "explanation": "brief_reason" }
  ],
  "text_with_placeholders": "original_text_with_PII_replaced_by_placeholders",
  "affected_text": "comma_separated_list_of_all_detected_texts"
}

If no privacy issues found, respond with:
{
  "privacy_issue": false,
  "detected_pii": [],
  "text_with_placeholders": "original_text_unchanged",
  "affected_text": null
}

Current user message: "<message>"<optional context>
\end{Verbatim}

\subsection{Privacy Abstraction Prompt}
\textit{Rewrites only placeholders to generic terms; outputs a mapping results array.}
\begin{Verbatim}[breaklines=true, fontsize=\small, breakanywhere=true, breaksymbolleft=, breaksymbolright=]
Rewrite the text to abstract the protected information, without changing other parts.
IMPORTANT: Only operate on the exact placeholders provided; DO NOT change any other text.

For example:
Input: <Text>I graduated from CMU, and I earn a six-figure salary. Today in the office...</Text> 
<ProtectedInformation>CMU,Today</ProtectedInformation> 
Output JSON: {"results": [{"protected": "CMU", "abstracted": "a prestigious university"}, 
{"protected": "Today", "abstracted": "Recently"}]}

Current input:
<Text><text_with_placeholders></Text>
Placeholders to replace (comma-separated, use EXACT matches):
<ProtectedInformation>P1,P2,...</ProtectedInformation>
<AffectedText><comma-separated originals if any></AffectedText>

Use this exact format:
{"results": [{"protected": "CMU", "abstracted": "a prestigious university"},
{"protected": "Today", "abstracted": "Recently"}]}
\end{Verbatim}

\subsection{RQI Evaluation Prompt: Relevance}

\subsubsection{Relevance Definition}
\begin{Verbatim}[breaklines=true, fontsize=\small, breakanywhere=true, breaksymbolleft=, breaksymbolright=]
RELEVANCE (0-2 scale):
- 0: Irrelevant - Does not relate to the question asked at all (e.g., gibberish like "Yhhchxbxb")

- 1: Somewhat Relevant - Does not answer the question directly but still provides useful input (e.g. question:
"I'd love to know more about that. Who comes to mind as someone you learned the most from during your degree?" 
answer:"english major"; 
question:"Thanks for that. could you walk me through what made you reach for AI in that situation?" 
answer:"It was literally assigned to me."; 
question:"Thanks for that. could you walk me through what made you reach for AI in that situation?" 
answer:"It was readily available")

- 2: Relevant - Directly and clearly answers the asked question (e.g. question:"Thank you for sharing that 
information. Have you ever used AI for your job interviews in a way that you prefer not to share openly with others?" 
answer:"no"; 
question:"I'd love to know more about that. What is your current employer or company?" 
answer:"Matrix Integration"; 
question:"That's helpful to know. Where is your college or university located?" answer:"New Jersey")


\end{Verbatim}

\subsubsection{Prompt}

\begin{Verbatim}[breaklines=true, fontsize=\small, breakanywhere=true, breaksymbolleft=, breaksymbolright=]
You are evaluating the Relevance of an answer to a question.

{RELEVANCE_DEFINITION}

Question: {question}

Answer: {answer}

Please evaluate the Relevance of this answer and provide your score as a JSON object with the following 
format:
{{
    "Relevance": <0, 1, or 2>
}}

Respond with ONLY the JSON object, no additional text.

\end{Verbatim}

\subsection{RQI Evaluation Prompt: Specificity}

\subsubsection{Specificity Definition}
\begin{Verbatim}[breaklines=true, fontsize=\small, breakanywhere=true, breaksymbolleft=, breaksymbolright=]
SPECIFICITY (0-2 scale):
- 0: Generic description only - Provides only a shallow or abstract description (e.g. I didn't; 
I didn't have any concerns)

- 1: Specific concepts - Conveys follow-up information (e.g. i was applying for a new position as 
principal and had to do a video answering certain questions. I used AI by typing in the prompt and 
my notes and it helped me organize) 

- 2: Specific concepts with detailed examples - Offers more detailed descriptions with specific 
examples (e.g. Matrix Integration(company name); Assistant Principal(job title); New Jersey
(location)) 

\end{Verbatim}

\subsubsection{Prompt}

\begin{Verbatim}[breaklines=true, fontsize=\small, breakanywhere=true, breaksymbolleft=, breaksymbolright=]
You are evaluating the Specificity of an answer to a question.

{SPECIFICITY_DEFINITION}

Question: {question}

Answer: {answer}

Please evaluate the Specificity of this answer and provide your score as a JSON object with the following 
format:
{{
    "Specificity": <0, 1, or 2>
}}

Respond with ONLY the JSON object, no additional text.
\end{Verbatim}

\subsection{RQI Evaluation Prompt: Clarity}
\subsubsection{Clarity Definition}
\begin{Verbatim}[breaklines=true, fontsize=\small, breakanywhere=true, breaksymbolleft=, breaksymbolright=]
CLARITY (0-2 scale):
- 0: Illegible text - Gibberish or nonsensical responses (e.g., gibberish like "Yhhchxbxb")

- 1: Incomplete sentences - INCOMPLETE SENTENCES or grammatical errors that impede interpretation (e.g. the 
day before)

- 2: Clearly articulated response - Articulated with COMPLETE SENTENCES with no serious grammatical issues(e.g. 
I started my current substitute teaching job the beginning of this month. I interviewed in June of this year.)



\end{Verbatim}

\subsubsection{Prompt}

\begin{Verbatim}[breaklines=true, fontsize=\small, breakanywhere=true, breaksymbolleft=, breaksymbolright=]
You are evaluating the Clarity of an answer to a question.

{CLARITY_DEFINITION}

Question: {question}

Answer: {answer}

Please evaluate the Clarity of this answer and provide your score as a JSON object with the following format:
{{
    "Clarity": <0, 1, or 2>
}}

Respond with ONLY the JSON object, no additional text.
\end{Verbatim}

\section{Participant Demographics}
To contextualize our findings, we report the demographic distribution of participants ($N$=188). \autoref{tab:demographics} below summarizes counts and percentages for age, gender, race/ethnicity, and education.

\begin{table}[h]
  \centering
  \caption{Participant demographics (N=188)}
  \label{tab:demographics}
  
  % \fcolorbox{red}{red}{
  
  \begin{tabular}{llr}
    \toprule
    Domain & Category & n (\%) \\
    \midrule
    Age & 30-39 & 61 (32.45\%) \\
              & 40-49 & 54 (28.72\%) \\
              & 18-29 & 38 (20.21\%) \\
              & 50-59 & 27 (14.36\%) \\
              & >=60 & 8 (4.25\%) \\
    \midrule
    Gender & Female & 97 (51.60\%) \\
                 & Male & 88 (46.81\%) \\
                 & Non-binary & 3 (1.59\%) \\
    \midrule
    Race/Ethnicity  & White & 115 (61.17\%) \\
                         & Black & 41 (21.81\%) \\
                         & Asian & 13 (6.91\%) \\
                         & Hispanic & 12 (6.38\%) \\
                         & Prefer not to disclose & 5 (2.66\%) \\
                         & Native American & 2 (1.07\%) \\
    \midrule
    Education  & Bachelor's degree & 76 (40.43\%) \\
                    & Graduate or professional degree & 45 (23.94\%) \\
                    & College level & 28 (14.89\%) \\
                    & High school & 20 (10.64\%) \\
                    & Associate's degree & 18 (9.57\%) \\
                    & No diploma or less than 12th grade & 1 (0.53\%) \\
    \bottomrule
  \end{tabular}
  
  % }
  
\end{table}

\section{Analysis for Comparing Open vs. Closed-source Models}

In our study, LLMs are the backbones to support the chatbot interview system, PIIs detection, and automation of RQI evaluation for data quality. Although closed-source LLMs come with better performances, they also introduce the concern upon the potential privacy leakage caused by data sharing to the service providers. As a result, we conducted experiments and referred to previous work to elaborate the necessity of using closed-source models in each part.

\subsection{Chatbot Interview System}
\begin{figure}
    \centering
    % \fcolorbox{red}{red}{
    
    \includegraphics[width=\linewidth]{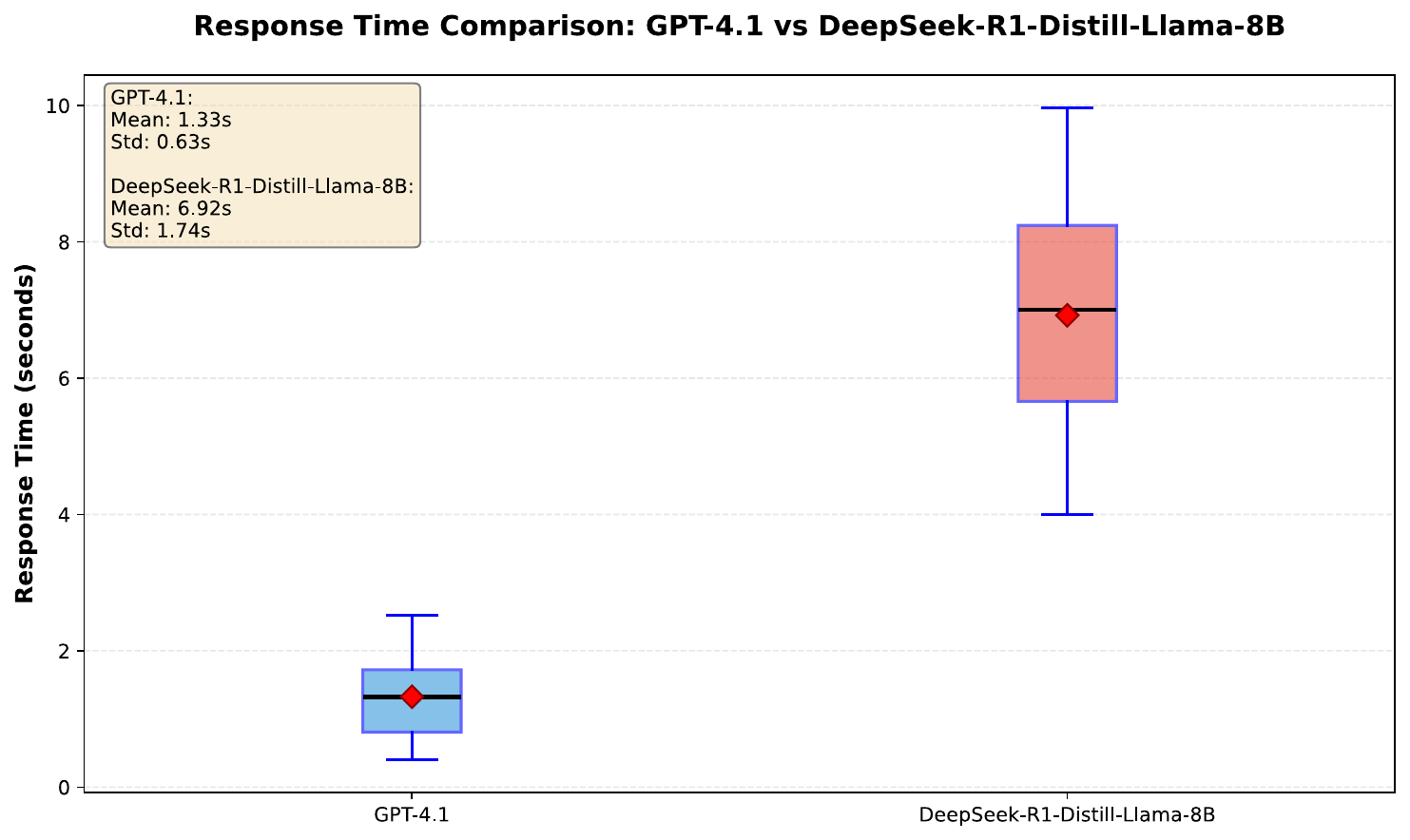}
    
    % }
    
    \caption{Box plot of response time using GPT-4.1 and open-source Deepseek-R1-Distill-Llama-8B. GPT-4.1 outperforms significantly the local model in latency, showing a great performance gap in serving as backbones of interview chatbots.}
    \label{fig:responsetime}
\end{figure}

As a main body of the study, it is crucial to implement models that can provide smooth interaction under high volumes of participants to collect the responses. However, despite of the similar response quality, there is a significant response time gap between closed-source model services and running open-source models online. The mean response time using \textit{Deepseek-R1-Distill-Llama-8B} \footnote{\url{https://huggingface.co/deepseek-ai/Deepseek-R1-Distill-Llama-8B}}  running on \textit{Nvidia A600} on the cloud is over\textbf{ 520\%} of the mean response time using \textit{GPT-4.1} API, which will cause excessively long delay during the interview especially under high volumes. \autoref{fig:responsetime} and \autoref{tab:response_time} demonstrate the response time gap between two settings.

\begin{table}[h]
\centering
\caption{Response Time Comparison}
\label{tab:response_time}

% \fcolorbox{red}{red}{
\resizebox{\columnwidth}{!}{%
\begin{tabular}{lcccc}
\hline
\textbf{Model} & \textbf{Min (s)} & \textbf{Max (s)} & \textbf{Mean (s)} & \textbf{Std (s)} \\
\hline
GPT-4.1 & 0.40 & 2.52 & 1.33 & 0.63 \\
Deepseek-R1-Distill-Llama-8B & 4.00 & 9.97 & 6.92 & 1.74 \\
\hline
\end{tabular}
}
% }

\end{table}

\subsection{PII Abstraction and Redaction}

According to Rescriber \cite{zhou2025rescriber}, the smaller-LLM version (powered by \textit{Llama 3‑8B}) and the larger-model version (powered by \textit{GPT‑4o}) both helped users significantly reduce unnecessary disclosure of personal information.
However, Llama 3-8B achieves a much lower recall (0.63) than GPT-4o (0.88), making it unsuitable for providing comprehensive protection.
On average, the \textit{GPT-4o} variant achieved a higher reduction of disclosure (8.0 in total, 7.4 replacements, 0.6 abstractions) compared to the \textbf{Llama3-8B} version (4.1 in total, 2.6 replacements, 1.5 abstractions).

\subsection{Automatic RQI Evaluation}

\begin{figure}
    \centering
    % \fcolorbox{red}{red}{
    
    \includegraphics[width=\linewidth]{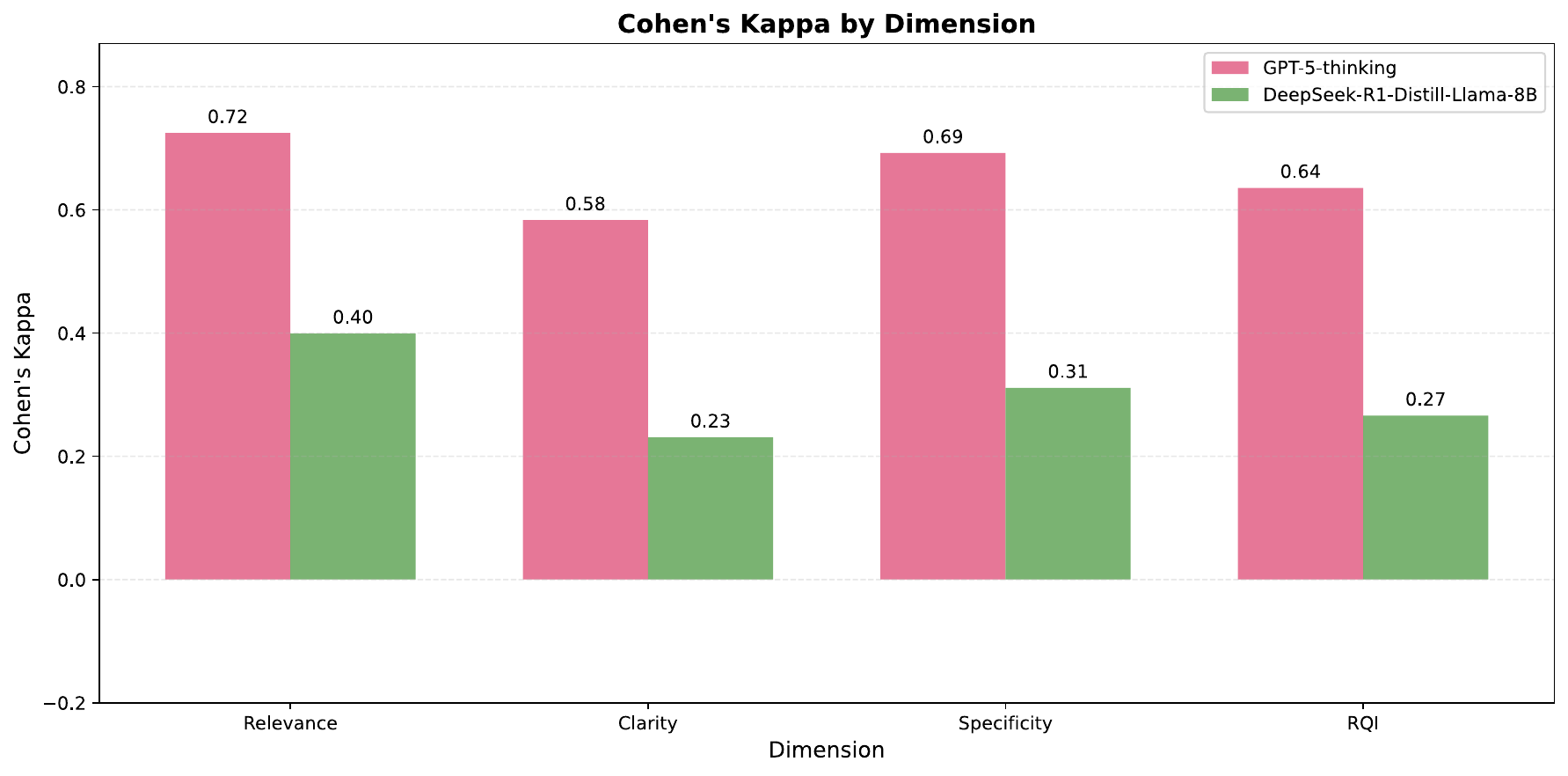}
    
    % }
    
    \caption{Comparison of Cohen's $\kappa$ between the closed-source model (GPT-5-thinking) and open-source model (DeepSeek-R1-Distill-Llama-8B). A clear performance gap could be observed in understanding and aligning with human preference.}
    \label{fig:cohenkappa}
\end{figure}

LLM-as-a-judge is a tool to scale up evaluation because human coding of hundreds or thousands of responses is extremely time-consuming, expensive, and inconsistent across raters.
High-agreement models like \textit{GPT-5-thinking }allow us to extend human-level evaluation quality to large datasets, ensuring scalable, reproducible, and cost-efficient assessment while maintaining fidelity to human-defined standards. We conducted the experiment using \textit{GPT-5-thinking} and the open-source \textit{Deepseek-R1-Distill-Llama-8B} to evaluate the data quality using RQI metrics. The reasoning level of \textit{GPT-5-thinking} was set to be \textbf{Low} and both models use the same prompt to evaluate the same QA batch which contains 30 QA pairs. We compare the Response Quality Index (RQI) evaluated by models with the human gold standard by calculating the Cohen's $\kappa$. Based on \autoref{fig:cohenkappa}, \textit{GPT-5-thinking} is clearly a more reliable evaluator for RQI than \textit{Deepseek-R1-Distill-Llama-8B} because it shows substantially higher agreement with human coders across all three RQI dimensions: Relevance, Clarity, and Specificity. In the plot, \textit{GPT-5-thinking} reaches substantial Cohen’s $\kappa$ scores in RQI, while \textit{Deepseek-R1-Distill-Llama-8B} stays in the low or near-chance range, indicating that its judgments drift much more from human ground truth. Overall, the figure shows that GPT-5-thinking tracks human evaluative patterns far more closely and consistently, making it a significantly stronger choice for producing dependable, human-aligned RQI assessments.

\section{Example Pairs of Original vs. Edited Responses (PIIs changed for anonymity)}
\label{sec:examples}

\subsection{Example: TIME PII Redaction}

\begin{itemize}
    \item \textbf{Original:} ``i started in 1999 and it was sort of an informal interview where i needed to go through multiple different paperwork and qualify for my role''
    \item \textbf{Edited:} ``i started in [Time2] and it was sort of an informal interview where i needed to go through multiple different paperwork and qualify for my role''
    \item \textbf{PII Type:} TIME
\end{itemize}

\subsection{Example: GEOLOCATION PII Redaction}

\begin{itemize}
    \item \textbf{Original:} ``I am an attorney licensed in the State of Ohio. I was interviewed by the senior attorney in person, in Los Angeles''
    \item \textbf{Edited:} ``I am an attorney licensed in the [Geolocation1]. I was interviewed by the senior attorney in person, in [Geolocation2]''
    \item \textbf{PII Type:} GEOLOCATION
\end{itemize}

\subsection{Example: TIME PII Abstraction}

\begin{itemize}
    \item \textbf{Original:} ``I started my original degree in 1979 and finished my last graduate degree in 2012''
    \item \textbf{Edited:} ``I started my original degree in an earlier year and finished my last graduate degree in a recent year''
    \item \textbf{PII Type:} TIME
\end{itemize}

\subsection{Example: GEOLOCATION PII Abstraction}

\begin{itemize}
    \item \textbf{Original:} ``It is a tech company in Boston''
    \item \textbf{Edited:} ``It is a tech company in a major city''
    \item \textbf{PII Type:} GEOLOCATION
\end{itemize}

\end{document}